\documentclass[prb,reprint,a4paper,showpacs,citeautoscript,floatfix]{revtex4-1}

\pdfoutput=1
\usepackage[charter]{mathdesign}
\usepackage{amsmath}
\usepackage[pdftex]{graphicx}
\usepackage{microtype}
\usepackage{bm}\let\vec\bm

\usepackage[svgnames]{xcolor}
\usepackage[
	colorlinks=True,linkcolor=DarkRed,citecolor=ForestGreen,urlcolor=MediumBlue,
	pdfstartview=FitH,bookmarks=False,pdfpagemode=UseNone
]{hyperref}

\begin{document}

\title{Bogoliubov quasiparticles coupled to the antiferromagnetic spin mode in a vortex core}

\author{C. Berthod}
\affiliation{Department of Quantum Matter Physics, University of Geneva, 24 quai Ernest-Ansermet, 1211 Geneva, Switzerland}

\date{August 19, 2015}

\begin{abstract}

In copper- and iron-based unconventional superconductors, the Bogoliubov quasiparticles interact with a spin resonance at momentum $(\pi,\pi)$. This interaction is revealed by specific signatures in the quasiparticle spectroscopies, like kinks in photoemission and dips in tunneling. We study these signatures, as they appear inside and around a vortex core in the local density of states (LDOS), a property accessible experimentally by scanning tunneling spectroscopy. Our model retains the whole nonlocal structure of the self-energy in space and time and is therefore not amenable to a Hamiltonian treatment using Bogoliubov--de Gennes equations. The interaction with the spin resonance does not suppress the zero-bias peak at the vortex center, although it reduces its spectral weight; neither does it smear out the vortex LDOS, but rather it adds structure to it. Some of the signatures we find may have been already measured in FeSe, but remained unnoticed. We compare the LDOS as a function of both energy and position with and without coupling to the spin resonance and observe, in particular, that the quasiparticle interference patterns around the vortex are strongly damped by the coupling. We study in detail the transfer of spectral weight induced both locally and globally by the interaction and also by the formation of the vortex. Finally, we introduce a new way of imaging the quasiparticles in real space, which combines locality and momentum-space sensitivity. This approach allows one to access quasiparticle properties that are not contained in the LDOS.

\end{abstract}

\pacs{74.55.+v, 74.72.-h, 74.20.Pq}
\maketitle

\section{Introduction}

A renewed interest in topological states of matter \cite{Hasan-2010} has contributed to bring again into focus the zero modes, which appear at the interface between regions with mismatched topologies. The electronic states bound to the core of vortices in type-II superconductors belong to this family \cite{Volovik-2003, Roy-2010, Herbut-2010}. They are tied to the nontrivial topology of the vortices, which carry an odd winding number of the order-parameter phase. These states---which are rather near-zero modes, because the topology may not impose a state at \emph{exactly} zero energy in all cases---were predicted by the Bogoliubov--de Gennes equations \cite{Caroli-1964, Gygi-1991} and directly observed experimentally in NbSe$_2$ by scanning tunneling spectroscopy (STS), providing a striking verification of the theory \cite{Hess-1989, Guillamon-2008a}. In cuprate high-$T_c$ superconductors (HTS), similar STS experiments in vortex cores \cite{Maggio-Aprile-1995, Renner-1998b, Hoogenboom-2000a, Pan-2000b, Matsuba-2003a, *Matsuba-2007, Shibata-2003b, *Shibata-2010, Levy-2005, Yoshizawa-2013} failed to reveal the signature expected for bound states in a $d$-wave superconductor \cite{Wang-1995, Franz-1998b, Yasui-1999}. This is surprising, because zero modes are in principle protected by topology and should be robust. Several explanations have been put forward \cite{Arovas-1997, Himeda-1997, Andersen-2000, Wu-2000, Kishine-2001, Berthod-2001b, Zhu-2001a, *Chen-2002b, Maska-2003, Tsuchiura-2003, Takigawa-2003, *Takigawa-2004, Fogelstrom-2011}, but a definitive interpretation of the vortex-core tunneling spectrum in high-$T_c$ oxides remains to be found. By contrast, the iron-based high-$T_c$ superconductors generally present vortex cores with the expected zero-energy peak characteristic of the bound states \cite{Song-2011, Shan-2011, Hanaguri-2012}, although in one case there are core states but no peak.\cite{Yin-2009}

In dirty superconductors, the zero-bias peak associated with the bound states is broadened \cite{Renner-1991, Eskildsen-2002}, leading to a flat tunneling spectrum in the core. This alone cannot resolve the cuprate puzzle, because the vortex-core spectrum is \emph{gapped} at zero bias in these materials, with features reminiscent of the spectrum observed in the pseudogap phase above $T_c$ \cite{Fischer-2007}. This has lead to the idea that the vortex cores are electronically similar to the mysterious pseudogap phase. It remains unclear how  the pseudogap and superconductivity would interact in the vortex core and, in particular, how this interaction could release the topological frustration which demands a zero mode.

Far from vortices, or in zero field, the low-temperature tunneling spectrum of bismuth-based HTS is rather well understood. It was recently found that an extension of the BCS theory, taking into account the band structure and a coupling to the antiferromagnetic spin resonance \cite{Eschrig-2000}, is able to reproduce the STS data of Bi$_2$Sr$_2$Ca$_2$Cu$_3$O$_{10+\delta}$ (Bi-2223) quantitatively \cite{Berthod-2013}. This modeling shows that the interaction with the spin resonance changes the density of states significantly in zero field, by redistributing spectral weight over a broad energy range. A question naturally follows: how would this interaction change the electronic structure in a vortex? One possibility is that the antiferromagnetic order becomes static in the cores \cite{Arovas-1997}, as several experiments have suggested \cite{Vaknin-2000, Lake-2001, *Lake-2002, Kakuyanagi-2003, Mounce-2011}. The local density of states (LDOS) in a vortex core with competing antiferromagnetic order may indeed share some similarities with the STS data for the cuprates \cite{Zhu-2001a, Takigawa-2003, *Takigawa-2004}. In the present work, we explore the opposite scenario, in which the antiferromagnetic fluctuations are not frozen, but remain dynamical in the vortex core. In contrast to the static case, the dynamical case cannot be formulated as a Hamiltonian mean-field problem: the coupling to the spin resonance enters via a nonlocal and energy-dependent self-energy. A simplified version of this model, ignoring nonlocal terms in the self-energy, was used earlier to study the effect of the spin resonance on the LDOS around a nonmagnetic impurity \cite{Zhu-2004a}. Very recently, the same approach was applied in a small cluster to investigate charge-density wave formation \cite{Bauer-2015}. Here, we solve this problem for a vortex of $d_{x^2-y^2}$ symmetry in a two-dimensional one-band system with parameters appropriate for Bi-2223. Although our results for the LDOS disagree with the vortex-core measurements in this material,\footnote{N. Jenkins, private communication.} they show how the tunneling spectrum would look like in Bi-2223, in the absence of a pseudogap or on the strongly overdoped side, if the interpretation given in zero field in terms of Bogoliubov quasiparticles interacting with the spin resonance is correct. The spin fluctuations are a candidate for the pairing interaction in the iron-based superconductors, as they can generate the attraction for a pairing with $s^{\pm}$ symmetry between different Fermi-surface sheets in these multi-band systems. We believe that many of our results are relevant for the qualitative understanding of the recently measured vortex spectra in these materials. We will argue that the measurements reported in Ref.~\onlinecite{Song-2011} contain signatures of the interaction with the spin resonance.

The STS studies have so far been limited to---more precisely, \emph{designed} to---measuring locally, targeting the LDOS, which is the diagonal part of the real-space electron Green's function. Some important properties of the quasiparticles, such as their nodal or antinodal character or their mean free path, do not appear clearly in the LDOS, but only indirectly, for instance via quasiparticle interference. After having discussed the LDOS, we will present another way of imaging the quasiparticles directly in real space. This approach probes the off-diagonal terms of the Green's function and combines locality with nonlocal (momentum-space) sensitivity. Such a measurement is a considerable experimental challenge, requiring finely controlled double-tip tunneling, but could greatly enrich our knowledge of quasiparticles in correlated metals. We illustrate this by showing, in particular, that the zero-energy quasiparticles remain nodal in the vortex core, despite the fact that the zero-energy LDOS extends along the antinodal directions.

The model used and the methods employed to solve it are described in Sec.~\ref{sec:model} and Appendices~\ref{app:symmetry} and \ref{app:Chebyshev}. The results are presented in Sec.~\ref{sec:results}: the self-consistent vortex order parameter in Sec.~\ref{sec:self-consistent-gap}, the LDOS and a study of the spectral-weight transfer in Sec.~\ref{sec:LDOS}, and the new quasiparticle imaging in Sec.~\ref{sec:NLCR} and Appendix~\ref{app:double-tip}. A discussion is proposed in Sec.~\ref{sec:discussion}.

\section{Model and methods}\label{sec:model}

The Bogoliubov quasiparticles of a superconductor can emit or absorb spin fluctuations and thereby become short-lived, if they are coupled to the spectrum of spin excitations. In a translation-invariant superconductor, these inelastic processes are described by self-energy corrections in momentum space, which modify both the normal (quasiparticle) and anomalous (gap) dispersions. At leading order, the self-energy is proportional to the convolution of the spin susceptibility with the quasiparticle propagator \cite{Eschrig-2000}. In real space, the momentum convolution becomes a product and the self-energy takes the form
	\begin{multline}\label{eq:Sigma1}
		\hat{\Sigma}(\vec{r}-\vec{r}',i\omega_n)=-\frac{1}{\beta}\sum_{i\Omega_n}
		g^2\chi_s(\vec{r}-\vec{r}',i\Omega_n)\\ \times\hat{\mathscr{G}}_{\text{BCS}}
		(\vec{r}-\vec{r}',i\omega_n-i\Omega_n).
	\end{multline}
The self-energy is a matrix in Nambu space: the matrix element $\Sigma_{11}$ ($\Sigma_{22}$) describes the renormalization and lifetime of particlelike (holelike) quasiparticles and the matrix elements $\Sigma_{12}$ and $\Sigma_{21}$ contain the renormalization and lifetime effects for the superconducting gap. The symmetry relations connecting these matrix elements are discussed below. In Eq.~(\ref{eq:Sigma1}), the self-energy is a function of the fermionic Matsubara frequency $\omega_n=(2n+1)\pi/\beta$ with $\beta=(k_{\mathrm{B}}T)^{-1}$ the inverse temperature, while the spin susceptibility $\chi_s$ is a function of the bosonic frequency $\Omega_n=2n\pi/\beta$. $\hat{\mathscr{G}}_{\mathrm{BCS}}$ is the Green's function describing the BCS--Bogoliubov quasiparticles in Nambu space, in the absence of coupling to the spin excitations. Finally, $g$ is a coupling parameter. The justification for a constant (momentum-independent) coupling is that for the cuprates, in the energy range of interest, the spin susceptibility is dominated by the antiferromagnetic resonance \cite{Eschrig-2006}, the weight of which is mostly localized near the momentum $(\pi,\pi)$.

Since the spin susceptibility has a sharp structure at the spin-resonance energy $\Omega_s$, while the BCS Green's function has structure at the gap edges $\sim\Delta$, the main structure of the self-energy develops around $\Omega_s+\Delta$, producing a kink in the quasiparticle dispersion and a dip in the tunneling spectrum \cite{Eschrig-2000}. The interplay of the van Hove singularity can induce additional structures and change these energies slightly \cite{Levy-2008, Berthod-2013}. In the core of a BCS vortex, the main structure of the Green's function comes from the bound states near zero energy. One may therefore expect that the scattering rate due to spin fluctuations is largest at the energy $\Omega_s$ in the core, which would produce a dip at this energy in the final vortex-core spectrum. This naive expectation may miss part of the story, however, because it assumes a purely local effect of spin fluctuations, while in the homogeneous case the self-energy has a marked momentum dependence \cite{Eschrig-2000, Berthod-2010}, indicating significant nonlocal components.

\begin{figure}[tb]
\includegraphics[width=\columnwidth]{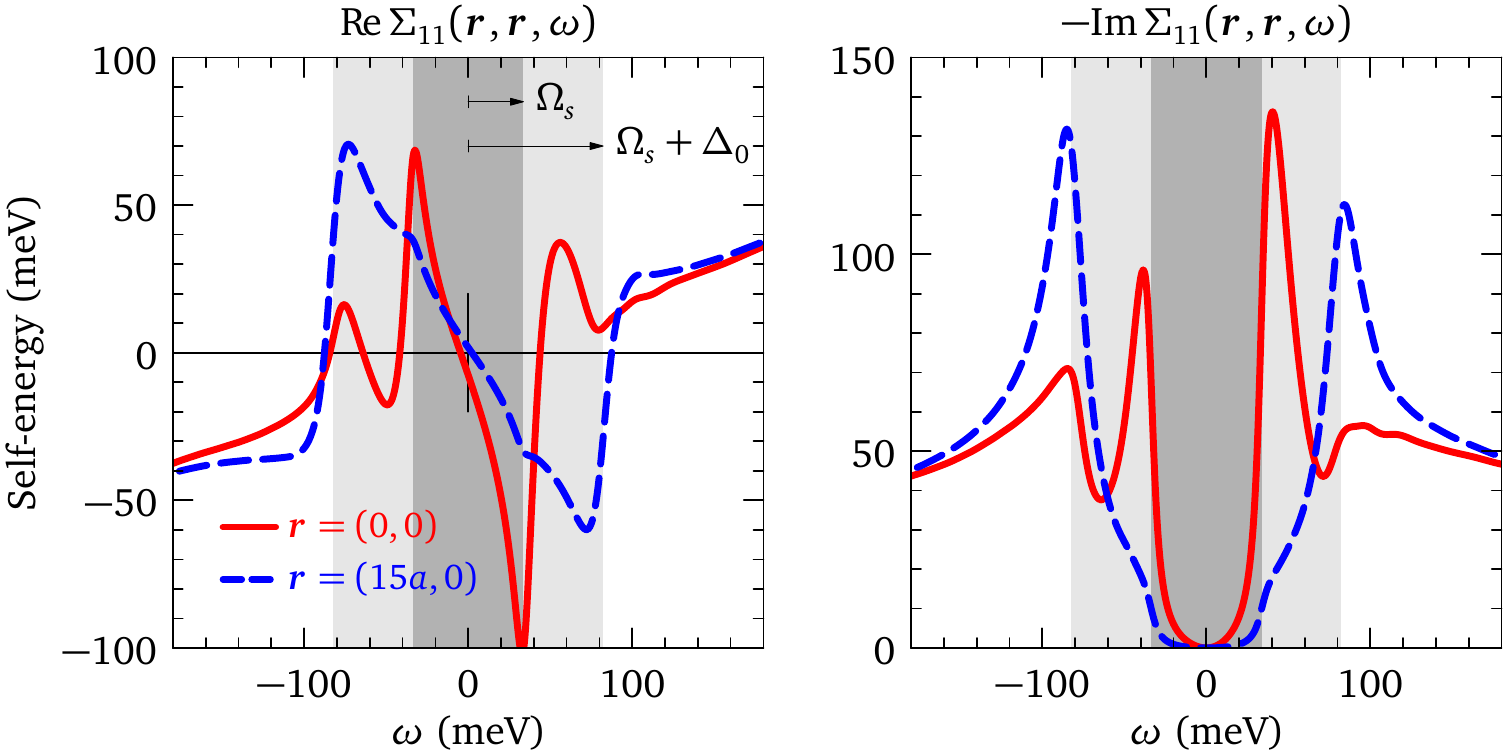}
\caption{\label{fig:fig-Sigma}
Local self-energy at the center of an isolated vortex (solid red lines) and outside the vortex (dashed blue lines) for electronlike quasiparticles coupled to a $(\pi,\pi)$ mode of energy $\Omega_s$ at $T=2$~K. The dark gray regions indicates energies $|\omega|<\Omega_s$ and the light gray regions correspond to $\Omega_s<|\omega|<\Omega_s+\Delta_0$. All model parameters are the same as in Sec.~\ref{sec:results}.
}
\end{figure}

In order to compute the effect of spin fluctuations on the tunneling spectrum in a $d$-wave vortex, we replace the Green's function $\hat{\mathscr{G}}_{\mathrm{BCS}}$ of a uniform BCS $d$-wave superconductor in expression (\ref{eq:Sigma1}) by the BCS Green's function $\hat{\mathscr{G}}_{\mathrm{vtx}}$ calculated in the presence of a vortex. The latter is significantly modified with respect to $\hat{\mathscr{G}}_{\mathrm{BCS}}$ along with the formation of the core states and, in particular, looses translation invariance:
	\begin{multline}\label{eq:Sigma2}
		\hat{\Sigma}(\vec{r},\vec{r}',i\omega_n)=-\frac{1}{\beta}\sum_{i\Omega_n}
		g^2\chi_s(\vec{r}-\vec{r}',i\Omega_n)\\ \times\hat{\mathscr{G}}_{\text{vtx}}
		(\vec{r},\vec{r}',i\omega_n-i\Omega_n).
	\end{multline}
We neglect a possible feedback of the vortices on the spin susceptibility and assume that it remains translation invariant. We shall compute the self-energy (\ref{eq:Sigma2}) numerically on the real-frequency axis, as explained further below. An example of a numerical result is shown in Fig.~\ref{fig:fig-Sigma}. With the self-energy ready, the last step in order to obtain the vortex-core spectrum is to solve the modified Gorkov equations, written hereafter in matrix form with the Nambu indices explicit and for an arbitrary complex energy $z$:
	\begin{multline}\label{eq:Gorkov}
		\begin{pmatrix}\mathscr{G}_0^{-1}(z)-\Sigma_{11}(z)&-\Delta-\Sigma_{12}(z)\\
		-\Delta^{\dagger}-\Sigma_{21}(z) & -[\mathscr{G}_0^{-1}(-z)]^T-\Sigma_{22}(z)
		\end{pmatrix}\\ \times
		\begin{pmatrix}\mathscr{G}_{11}(z)&\mathscr{G}_{12}(z)\\ \mathscr{G}_{21}(z)&
		\mathscr{G}_{22}(z)\end{pmatrix}=\begin{pmatrix}\openone&0\\0&\openone\end{pmatrix}.
	\end{multline}
All products are matrix products in the implied spatial coordinates. $\mathscr{G}_0^{-1}(z)\equiv\mathscr{G}_0^{-1}(\vec{r},\vec{r}',z)=(z+\mu)\delta_{\vec{r}\vec{r}'}-t_{\vec{r}\vec{r}'}$ is the noninteracting normal-state Green's function, with $\mu$ the chemical potential and $t_{\vec{r}\vec{r}'}$ the hopping amplitude. \footnote{We use the bare hopping amplitudes and neglect a correction due to the magnetic field. This correction is negligible for an isolated vortex in the region of the core when the penetration depth is large compared to the core size.} $\Delta(\vec{r},\vec{r}')$ is the superconducting pair potential describing a vortex with $d_{x^2-y^2}$ symmetry. The symbol ``$T$'' means transposition of the spatial coordinates and ``$\dagger$'' means the same transposition followed by complex conjugation. It should be emphasized that the spin resonance is not the primary source of pairing in our model: the pair potential $\Delta(\vec{r},\vec{r}')$ is generated by a different interaction of unspecified origin, via static Bogoliubov--de Gennes equations. We are primarily interested in the component $\mathscr{G}_{11}$, whose diagonal elements provide the local density of states:
	\begin{equation}\label{eq:LDOS}
		N(\vec{r},\varepsilon)=-{\textstyle\frac{1}{\pi}}\text{Im}\,
		\mathscr{G}_{11}(\vec{r},\vec{r},z=\varepsilon+i\Gamma).
	\end{equation}
The Dyson equation for $\mathscr{G}_{11}$ is, as usual, $\mathscr{G}_{11}^{-1}=\mathscr{G}_0^{-1}-\Sigma$, with the corresponding self-energy obtained by solving Eq.~(\ref{eq:Gorkov}):
	\begin{multline}\label{eq:Sigma3}
		\Sigma(z)=\Sigma_{11}(z)-[\Delta+\Sigma_{12}(z)]\\
		\times \{[\mathscr{G}_0^{-1}(-z)]^T+\Sigma_{22}(z)\}^{-1}
		[\Delta^{\dagger}+\Sigma_{21}(z)].
	\end{multline}
The difficulty of this problem stems from the combination of broken translational invariance and nonlocality in time. The rest of this section is mostly technical and describes our practical implementation of the solution. An overview of the calculation workflow is given at the end of the section.

If the components $\Sigma_{ij}$ of (\ref{eq:Sigma2}) are given, the calculation of the self-energy (\ref{eq:Sigma3}) for one particular energy requires a matrix inversion in the spatial indices. The quantity $\{\cdots\}^{-1}$, which is the renormalized propagator of the holelike quasiparticles, is indeed better written as $(\openone-\mathscr{G}_0^T\Sigma_{22})^{-1}\mathscr{G}_0^T$. For practical reasons, such matrix inversions limit the system size to a few thousands lattice sites. One further matrix inversion is needed in order to obtain the local density of states, by solving the Dyson equation, also rewritten in the form $\mathscr{G}_{11}=(\openone-\mathscr{G}_0\Sigma)^{-1}\mathscr{G}_0$. The reason for preferring these rewritings, as opposed to simply solving, e.g., $\mathscr{G}_{11}=(\mathscr{G}_0^{-1}-\Sigma)^{-1}$---which also requires only one matrix inversion because $\mathscr{G}_0^{-1}$ is known analytically---is that they allow for a better energy resolution by minimizing finite-size effects. Thanks to its translation invariance, the matrix $\mathscr{G}_0$ is best computed as a Fourier transform $\mathscr{G}_0(\vec{r},\vec{r}',z)=(1/N)\sum_{\vec{k}}e^{i\vec{k}\cdot(\vec{r}-\vec{r}')}/(z-\xi_{\vec{k}})$ with $\xi_{\vec{k}}$ the dispersion measured from the chemical potential. We do this on a dense mesh of $\vec{k}$ vectors, with $N=1024\times1024$ points in the two-dimensional Brillouin zone, such that boundary effects are negligible. The expression $(\openone-\mathscr{G}_0\Sigma)^{-1}\mathscr{G}_0$ thus delivers a Green's function free of boundary effects if $\Sigma=0$, as opposed to the expression $(\mathscr{G}_0^{-1}-\Sigma)^{-1}$.

Let us now turn to the practical calculation of Eq.~(\ref{eq:Sigma2}) for real frequencies. Following previous studies \cite{Eschrig-2000, Levy-2008, Berthod-2013}, we use for the spin susceptibility a phenomenological model inspired by experiments:
	\begin{equation}\label{eq:chis}
		\chi_s(\vec{r}-\vec{r}',i\Omega_n)=W_sF(\vec{r}-\vec{r}')
		\int_{-\infty}^{\infty} d\varepsilon\,\frac{I(\varepsilon)}{i\Omega_n-\varepsilon}.
	\end{equation}
$I(\varepsilon)$ represents a Lorentzian-broadened spin resonance at frequency $\Omega_s$ and $F(\vec{q})$ is peaked around the antiferromagnetic vector $(\pi,\pi)$. In order to evaluate analytically the frequency sum in Eq.~(\ref{eq:Sigma2}), we use the representation of the Matsubara Green's function in terms of retarded (R) and advanced (A) functions
	\begin{align}
		\nonumber
		\hat{\mathscr{G}}_{\text{vtx}}(\vec{r},\vec{r}',i\omega_n)&=\frac{i}{2\pi}
		\int_{-\infty}^{\infty}d\varepsilon\,\frac{\hat{G}^R_{\text{vtx}}
		(\vec{r},\vec{r}',\varepsilon)-\hat{G}^A_{\text{vtx}}(\vec{r},\vec{r}',\varepsilon)}
		{i\omega_n-\varepsilon}\\
		&\equiv\int_{-\infty}^{\infty}d\varepsilon\,\frac{\hat{\rho}
		(\vec{r},\vec{r}',\varepsilon)}{i\omega_n-\varepsilon}.
	\end{align}
Performing the frequency sum and the analytic continuation $i\omega_n\to\omega+i0^+$, we obtain the retarded self-energy on the real axis:
	\begin{align}
		\nonumber
		\hat{\Sigma}(\vec{r},\vec{r}',\omega)&=g^2W_sF(\vec{r}-\vec{r}')
		\int_{-\infty}^{\infty}d\varepsilon\,\hat{\rho}(\vec{r},\vec{r}',\varepsilon)\\
		\nonumber
		&\quad\times\int_{-\infty}^{\infty}d\varepsilon'\,I(\varepsilon')
		\frac{f(-\varepsilon)+b(\varepsilon')}{\omega-\varepsilon-\varepsilon'+i0^+}\\
		\label{eq:Sigma4}
		&=\alpha^2F(\vec{r}-\vec{r}')\int_{-\infty}^{\infty}d\varepsilon\,
		\hat{\rho}(\vec{r},\vec{r}',\varepsilon)B_0(\omega,\varepsilon),\\
		\label{eq:B0sym}
		B_0(\omega,E)&=-B_0^*(-\omega,-E)\\
		\nonumber
		&=\Lambda^2\int_{-\infty}^{\infty}d\varepsilon\,
		[L_{\Gamma_s}(\varepsilon-\Omega_s)-L_{\Gamma_s}(\varepsilon+\Omega_s)]\\
		&\quad\times\frac{f(-E)+b(\varepsilon)}{\omega-E-\varepsilon+i0^+}.
	\end{align}
The parameters $\alpha^2$ and $\Lambda$ are a dimensionless coupling and a typical energy scale (nearest-neighbor hopping), respectively (see Ref.~\onlinecite{Berthod-2013}). $L_{\Gamma_s}$ is the Lorentzian, with $\Gamma_s$ the energy width of the spin resonance, while $f$ and $b$ are the Fermi-Dirac and Bose-Einstein distribution functions, respectively. The function $B_0(\omega,E)$ is analytically known and corresponds to the function $B(\omega,E)$ of Ref.~\onlinecite{Berthod-2013}, in which the Dynes parameter $\Gamma$ is set to $0^+$. This function is peaked near $E=\omega-\Omega_s\mathrm{sign}(E)$, such that the energy integration in Eq.~(\ref{eq:Sigma4}) is well convergent. The Dynes broadening is absent from the function $B_0$, because it is already implemented in the vortex Green's functions $\hat{G}^{R,\,A}_{\text{vtx}}$.

We show now that the spectral functions $\rho_{ij}(\vec{r},\vec{r}',\varepsilon)$, needed for the calculation of the self-energy (\ref{eq:Sigma4}), can all be deduced from the element `$11$' of the retarded Nambu matrix $\hat{G}^R_{\text{vtx}}$, element `$11$' denoted hereafter simply $G_{\text{vtx}}$. The vortex matrix Green's function for a general complex energy $z$ is the solution of Eq.~(\ref{eq:Gorkov}) with $\Sigma_{ij}\equiv0$. The element `$11$' satisfies the Dyson equation $\mathscr{G}_{11}^{-1}=\mathscr{G}_0^{-1}-\Sigma_{\text{vtx}}$, with the BCS vortex self-energy given by setting $\Sigma_{ij}=0$ in Eq.~(\ref{eq:Sigma3}): $\Sigma_{\text{vtx}}=-\Delta\mathscr{G}_0^T(-z)\Delta^{\dagger}$. For real frequencies, we have $\mathscr{G}_{11}=G_{\text{vtx}}$ and
	\begin{multline}\label{eq:Sigmavtx}
		\Sigma_{\text{vtx}}(\vec{r},\vec{r}',\omega)=\\-\sum_{\vec{r}_1\vec{r}_2}
		\Delta(\vec{r},\vec{r}_1)\mathscr{G}_0(\vec{r}_2-\vec{r}_1,-\omega-i\Gamma)
		\Delta^*(\vec{r}',\vec{r}_2).
	\end{multline}
One can see (Appendix~\ref{app:symmetry}) that the solution of Eq.~(\ref{eq:Gorkov}) with $\Sigma_{ij}\equiv0$ has the analytic property
	$
		\mathscr{G}_{11}(z)=\mathscr{G}_{11}^{\dagger}(z^*),
	$
from which we deduce that $G_{11}^A(\varepsilon)=[G_{11}^R(\varepsilon)]^{\dagger}$ and further that
	\begin{align}\label{eq:rho11}
		\rho_{11}(\vec{r},\vec{r}',\varepsilon)&=\frac{i}{2\pi}\left[G_{\text{vtx}}
		(\vec{r},\vec{r}',\varepsilon)-G^*_{\text{vtx}}(\vec{r}',\vec{r},\varepsilon)\right]\\
		\nonumber
		&=\rho_{11}^*(\vec{r}',\vec{r},\varepsilon).
	\end{align}
Note that $G_{\text{vtx}}(\vec{r},\vec{r}',\varepsilon)\neq G_{\text{vtx}}(\vec{r}',\vec{r},\varepsilon)$, such that the spectral function $\rho_{11}$ is complex. We also have the property
	$
		\mathscr{G}_{22}(z)=-\mathscr{G}_{11}^*(-z^*),
	$
which implies
	\begin{align}\label{eq:rho22}
		\rho_{22}(\vec{r},\vec{r}',\varepsilon)&=\frac{i}{2\pi}\left[G_{\text{vtx}}
		(\vec{r}',\vec{r},-\varepsilon)-G_{\text{vtx}}^*(\vec{r},\vec{r}',-\varepsilon)\right]\\
		\nonumber
		&=\rho_{11}^*(\vec{r},\vec{r}',-\varepsilon).
	\end{align}
Together with the property (\ref{eq:B0sym}), the relations (\ref{eq:rho11}) and (\ref{eq:rho22}) imply that the particlelike and holelike self-energies are related by $\Sigma_{22}(\vec{r},\vec{r}',\omega)=-\Sigma_{11}^*(\vec{r},\vec{r}',-\omega)$. Finally, we have
	$
		\mathscr{G}_{12}(z)=\mathscr{G}_{21}^{\dagger}(z^*)=\mathscr{G}_{21}^*(-z^*)
	$
and we deduce
	\begin{align}\label{eq:rho21}
		\rho_{21}(\vec{r},\vec{r}',\varepsilon)&=\frac{i}{2\pi}[F^{+}_{\text{vtx}}
		(\vec{r},\vec{r}',\varepsilon)-F^{+}_{\text{vtx}}(\vec{r}',\vec{r},-\varepsilon)]\\
		\nonumber
		&=-\rho_{21}(\vec{r}',\vec{r},-\varepsilon)=\rho_{12}^*(\vec{r}',\vec{r},\varepsilon),
	\end{align}
where, following the usual notation, the anomalous vortex Green's function is
	$
		F_{\text{vtx}}^{+}(\varepsilon)\equiv\mathscr{G}_{21}(\varepsilon+i0^+).
	$
The latter can also be expressed in terms of $G_{\text{vtx}}$. Eq.~(\ref{eq:Gorkov}) with $\Sigma_{ij}\equiv0$ gives $\mathscr{G}_{21}(z)=-\mathscr{G}_0^T(-z)\Delta^{\dagger}\mathscr{G}_{11}(z)$, which means on the real axis:
	\begin{multline}\label{eq:Fdagger}
		F_{\text{vtx}}^{+}(\vec{r},\vec{r}',\varepsilon)=\\-\sum_{\vec{r}_1\vec{r}_2}
		\mathscr{G}_0(\vec{r}_1-\vec{r},-\varepsilon-i\Gamma)
		\Delta^*(\vec{r}_2,\vec{r}_1)G_{\text{vtx}}(\vec{r}_2,\vec{r}',\varepsilon).
	\end{multline}
The relations (\ref{eq:rho21}) impose a connection between the pairing self-energies, namely $\Sigma_{12}(\vec{r},\vec{r}',\omega)=\Sigma_{21}^*(\vec{r},\vec{r}',-\omega)$.\\

The numerical calculation runs as follows.
\begin{enumerate}
\item Set the model parameters (dispersion $\xi_{\vec{k}}$, pairing interaction) and determine the self-consistent BCS vortex pair potential $\Delta(\vec{r},\vec{r}')$; for this step, we use the Chebyshev expansion method described in Ref.~\onlinecite{Covaci-2010} and briefly recalled in Appendix~\ref{app:Chebyshev}.
\item Choose the system size and use Eq.~(\ref{eq:Sigmavtx}) and the corresponding Dyson equation to calculate the vortex Green's function $G_{\text{vtx}}$ and to deduce $F_{\text{vtx}}^{+}$ with Eq.~(\ref{eq:Fdagger}).
\item Calculate and store the spectral functions $\rho_{ij}$ using Eqs.~(\ref{eq:rho11}--\ref{eq:rho21}) for all relevant energies and positions.
\item Set the spin-resonance parameters, perform the energy integration in Eq.~(\ref{eq:Sigma4}) and store the self-energies $\Sigma_{ij}$.
\item Evaluate the modified vortex self-energy (\ref{eq:Sigma3}) on the real axis and solve the corresponding Dyson equation to obtain the Green's function and finally the LDOS from Eq.~(\ref{eq:LDOS}).
\end{enumerate}
All calculations of the self-energy (\ref{eq:Sigma3}) are performed at $T=2$~K, which is the base temperature of the experiments reported in Ref.~\onlinecite{Berthod-2013}. The self-consistent order parameter $\Delta(\vec{r},\vec{r}')$ is computed at $T=0$ for simplicity.

\section{Results}\label{sec:results}

\subsection{Self-consistent vortex pair potential}\label{sec:self-consistent-gap}

\begin{figure}[b]
\includegraphics[width=\columnwidth]{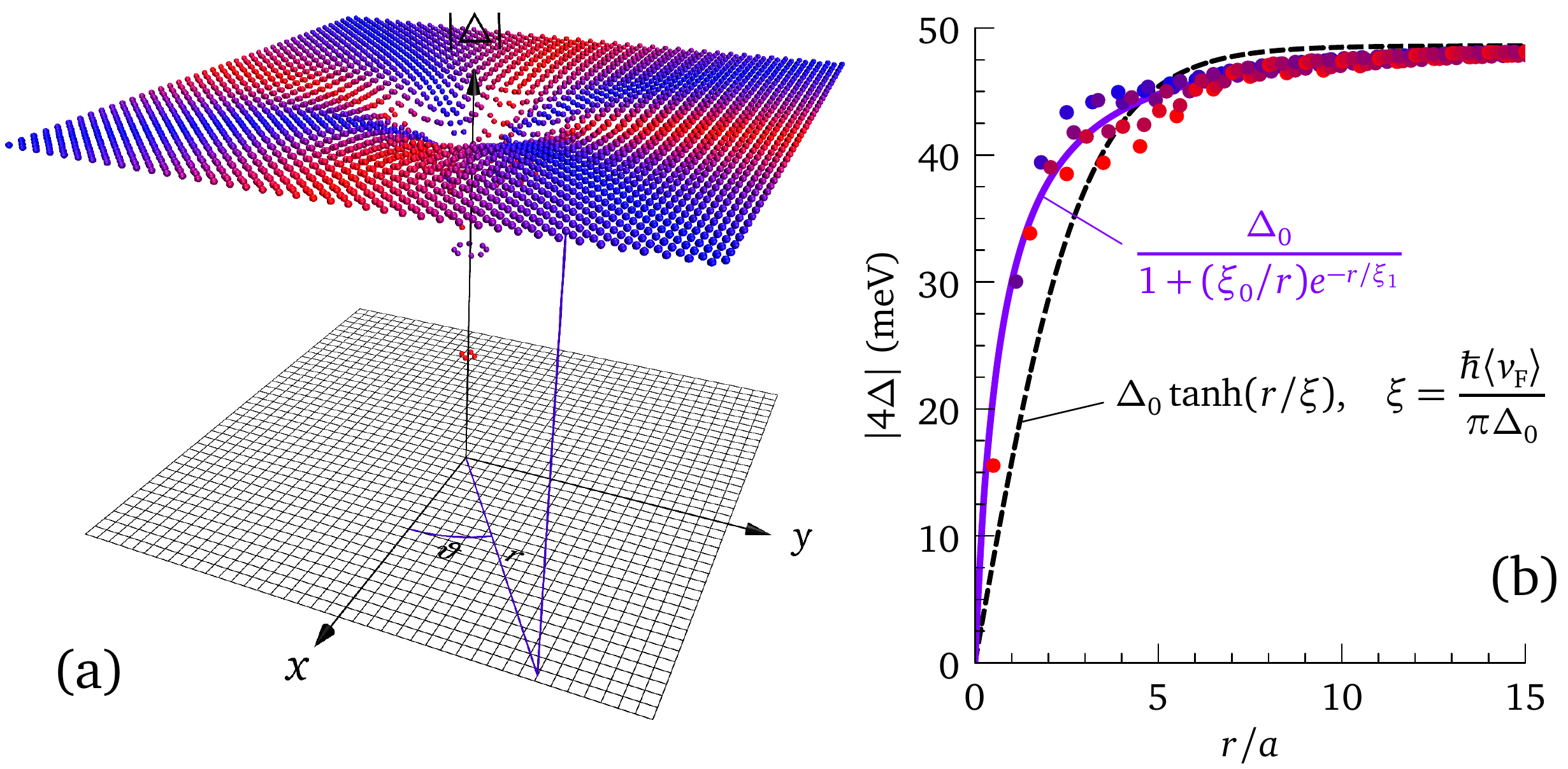}
\caption{\label{fig:fig-gap}
Self-consistent vortex pair potential calculated for the nearest-neighbor bonds on the square lattice, using the Chebyshev expansion (lattice size: $1001\times1001$, expansion order: 2000; relative convergence to self-consistency: $<6\times10^{-4}$; see Appendix~\ref{app:Chebyshev}). The color of dots in (a) and (b) varies from red (antinodal direction) to blue (nodal direction). The dashed black and solid purple lines in (b) show two functional dependencies as indicated, with $\xi=2.96a$, $\xi_0=0.68a$, and $\xi_1=10.5a$.
}
\end{figure}

A tight-binding model for the low-energy band structure of Bi-2223 was obtained in Ref.~\onlinecite{Berthod-2013}, by fitting tunneling spectra in zero field. We use the parameters corresponding to the spectrum with a peak-to-peak gap of 44~meV. These parameters are $(t_1,t_2,t_3,t_4,t_5,\mu)=(-206,56,-36,-10.3,27.9,-237)$~meV. The bare BCS $d$-wave gap is $\Delta_0=48.6$~meV. Figure~\ref{fig:fig-gap} displays the self-consistent BCS pair potential $\Delta(\vec{r},\vec{r}')$ calculated at $T=0$ for an isolated vortex. The pair potential vanishes in the core over a length scale similar to the bulk coherence length, consistently with previous Bogoliubov--de Gennes calculations for $d$-wave superconductors \cite{Soininen-1994, Wang-1995, Franz-1998b, Note3}. \footnotetext{In our model, the vortex center sits on a lattice site. We have also considered a vortex centered in the middle of a plaquette as in Ref.~\onlinecite{Soininen-1994}. Both are stable self-consistent solutions and yield nearly identical LDOS.} The pair potential is nonzero only on the nearest-neighbor bonds ($|\vec{r}-\vec{r}'|=a$). The pairing strength was adjusted to reproduce the bulk gap $\Delta_0$ far from the vortex. The dots in Fig.~\ref{fig:fig-gap}(a) show the modulus of the pair potential for each bond; this representation differs from that in Ref.~\onlinecite{Soininen-1994}, where local $d_{x^2-y^2}$ and extended-$s$ wave components were defined at each site. Beside the modulus shown in Fig.~\ref{fig:fig-gap}(a), the pair potential carries the $x^2-y^2$ signature, namely a plus (minus) sign on bonds running along $x$ ($y$), as well as the topological phase given to an excellent approximation by $-\vartheta$, where $\vartheta$ is the angle defined by the middle of the bond and the vortex center [see Fig.~\ref{fig:fig-gap}(a)].

The pair-potential modulus has cylindrical symmetry at large distances from the core, but presents some anisotropy at intermediate distances, as shown in Fig.~\ref{fig:fig-gap}(b). The gap increases faster along the $(1,1)$ direction than along the $(1,0)$ direction. By changing the band parameters, we have found that this behavior is model dependent. As already noted in Refs.~\onlinecite{Gygi-1991, Franz-1998b}, the relaxation of the gap is not very well described by the Ginzburg-Landau functional form, $\tanh(r/\xi)$, whatever the value of $\xi$. The BCS expression of the coherence length is $\xi=\hbar\langle v_{\mathrm{F}}\rangle/(\pi\Delta_0)$. With our band parameters, the average Fermi velocity is $2.63\times10^7$~cm/s and thus $\xi=1.13$~nm. The corresponding expected $\tanh(r/\xi)$ dependency is shown in Fig.~\ref{fig:fig-gap}(b) as the dashed line. No good fit can be achieved with any value of $\xi$. We find that a different two-parameter functional form fits the numerical data much better, as indicated by the purple line.

A self-consistent pair potential is nice to have, but not essential for the study of the LDOS in the core region \cite{Berthod-2005}. We have not found significant differences between the LDOS calculated using the Ginzburg-Landau and the self-consistent pair potentials. All data shown below correspond to the self-consistent case.

\subsection{Local density of states and spectral-weight transfer in the vortex coupled to the spin resonance}
\label{sec:LDOS}

\begin{figure}[tb]
\includegraphics[width=\columnwidth]{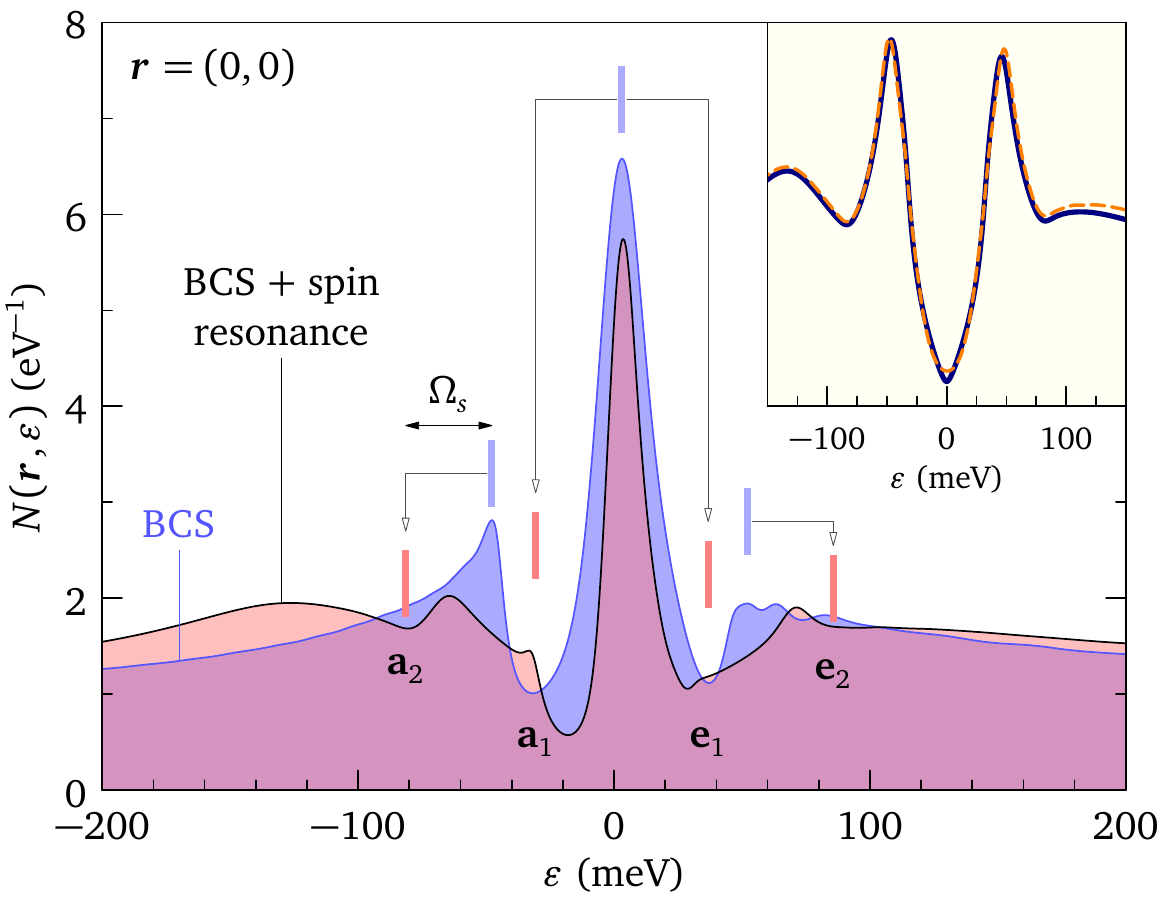}
\caption{\label{fig:fig-core}
Redistribution of spectral weight in the core of a $d$-wave vortex, due to interaction with an antiferromagnetic spin resonance of energy $\Omega_s$. The black line with pink shade is the LDOS with the interaction turned on, while for the shaded blue line (BCS) it is turned off. Vertical blue bars indicate the main spectral features of the BCS spectrum; pink bars indicate energies corresponding to the blue bars, shifted by $\pm\Omega_s$. (Inset) Illustration of finite-size effects. Solid line: converged DOS in zero field. Dashed line: LDOS at the central site, after replacing the vortex by a uniform $d$-wave gap.
}
\end{figure}

The spin resonance is characterized by its energy $\Omega_s=33.7$~meV, energy width $\Gamma_s=4.2$~meV, and momentum width $\Delta q=1.15/a$. With a coupling $\alpha=0.7$,\footnote{A coupling $\alpha=0.7$ in our notation corresponds, in the notation of Ref.~\onlinecite{Eschrig-2000}, to $g=0.77$~eV if the energy-integrated susceptibility at $(\pi,\pi)$ is $0.95\mu_{\mathrm{B}}^2$.} these parameters yield the zero-field DOS shown in the inset of Fig.~\ref{fig:fig-core}, in very good agreement with the experimental tunneling spectrum \cite{Berthod-2013}. Note that the peak maxima in this zero-field DOS define a gap $\Delta_p=46$~meV, slightly larger than the experimental gap of 44~meV. We calculate the vortex-core LDOS in a $71\times71$ cluster having the vortex at its center \cite{Note3}. In order to estimate the importance of finite-size effects, we replace the vortex pair potential by a uniform $d$-wave gap and compare the resulting LDOS at the central site with the fully converged DOS calculated in momentum space using $1024\times1024$ $\vec{k}$ points (inset of Fig.~\ref{fig:fig-core}). The good agreement between the two curves shows that the finite-size effects are small. Whenever this is possible, i.e., in the case without coupling to the spin mode, we also check our calculations against Chebyshev-expansion calculations in a much bigger cluster of size  $1001\times1001$ (Appendix~\ref{app:Chebyshev}).

\begin{figure}[tb]
\includegraphics[width=0.8\columnwidth]{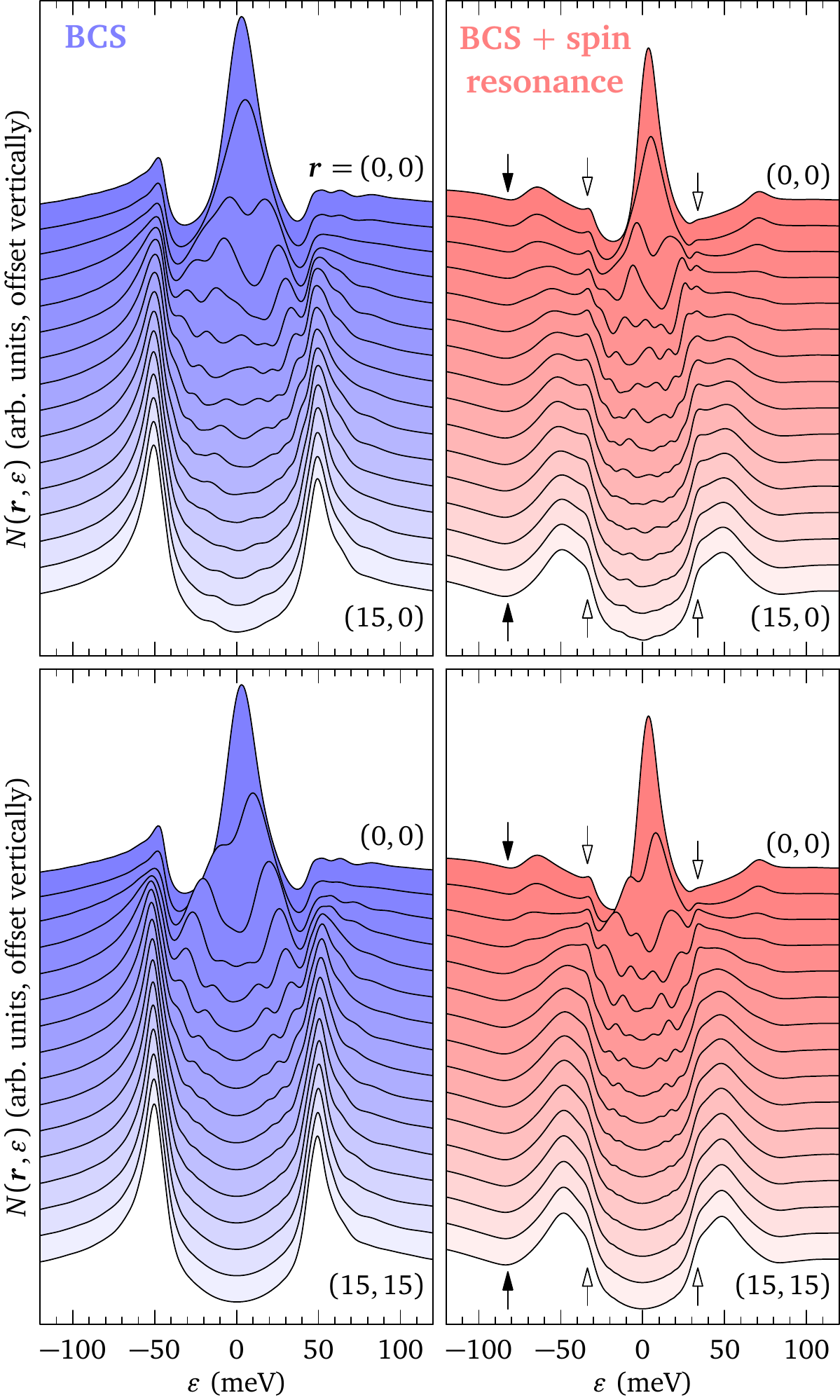}
\caption{\label{fig:fig-traces}
Comparison of the LDOS in the vicinity of the vortex core with the coupling to the spin resonance turned off (left) and on (right). Upper and lower panels show a trace starting at the core center and going along the $x$ axis and along the diagonal of the square lattice, respectively. The white and black arrows mark the energies $\pm\Omega_s$ and $-(\Omega_s+\Delta_0)$, respectively.
}
\end{figure}

\begin{figure}[tb]
\includegraphics[width=0.8\columnwidth]{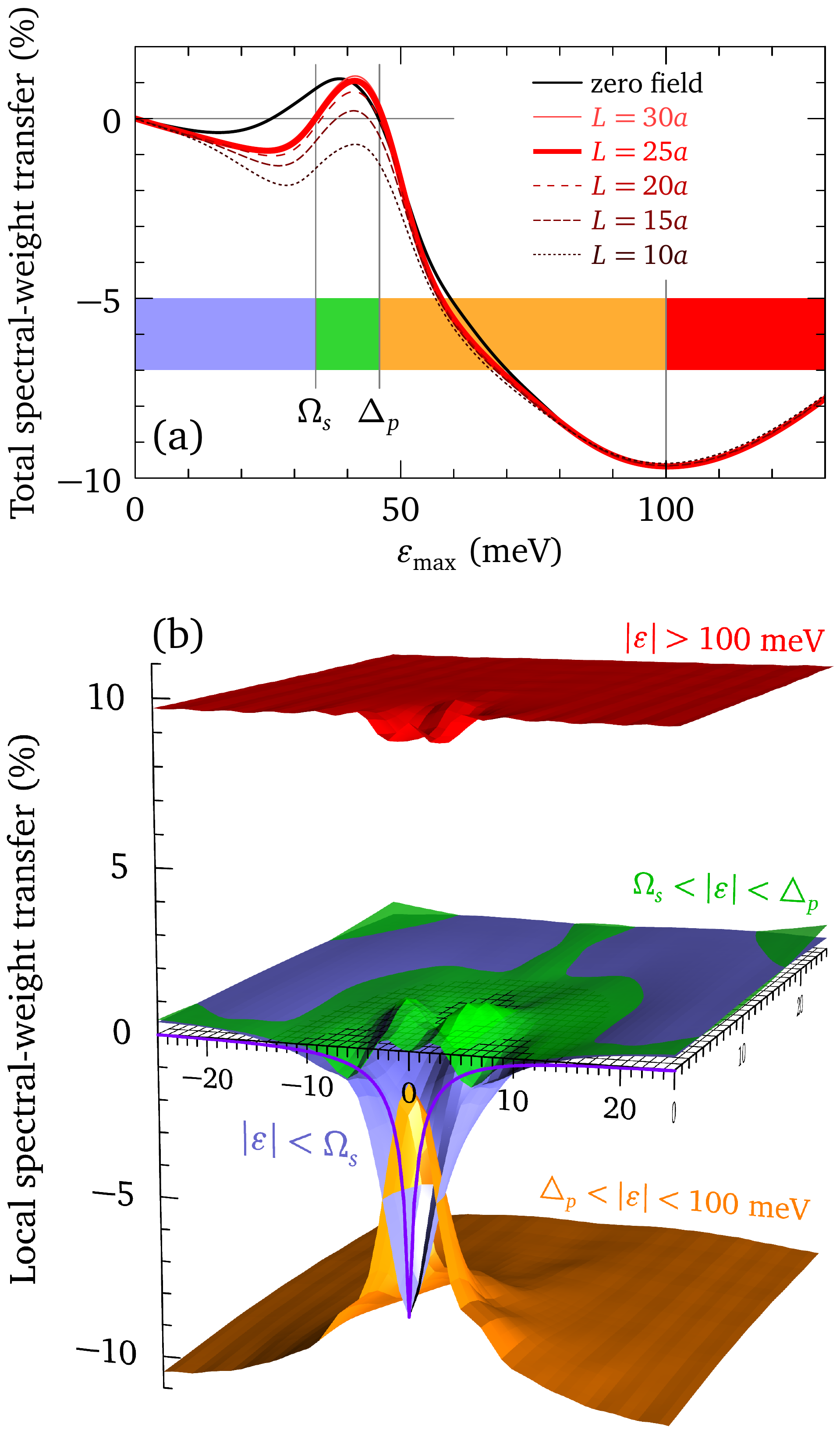}
\caption{\label{fig:fig-weight-1}
Total (a) and local (b) spectral-weight transfer induced by the spin resonance (see text). The colored ranges in (a) correspond to the four energy domains in (b). Only half of the spatial region defined by $L=25a$ is shown in (b). The purple line represents the modulus of the pair potential as determined in Fig.~\ref{fig:fig-gap}.
}
\end{figure}

\begin{figure}[tb]
\includegraphics[width=0.8\columnwidth]{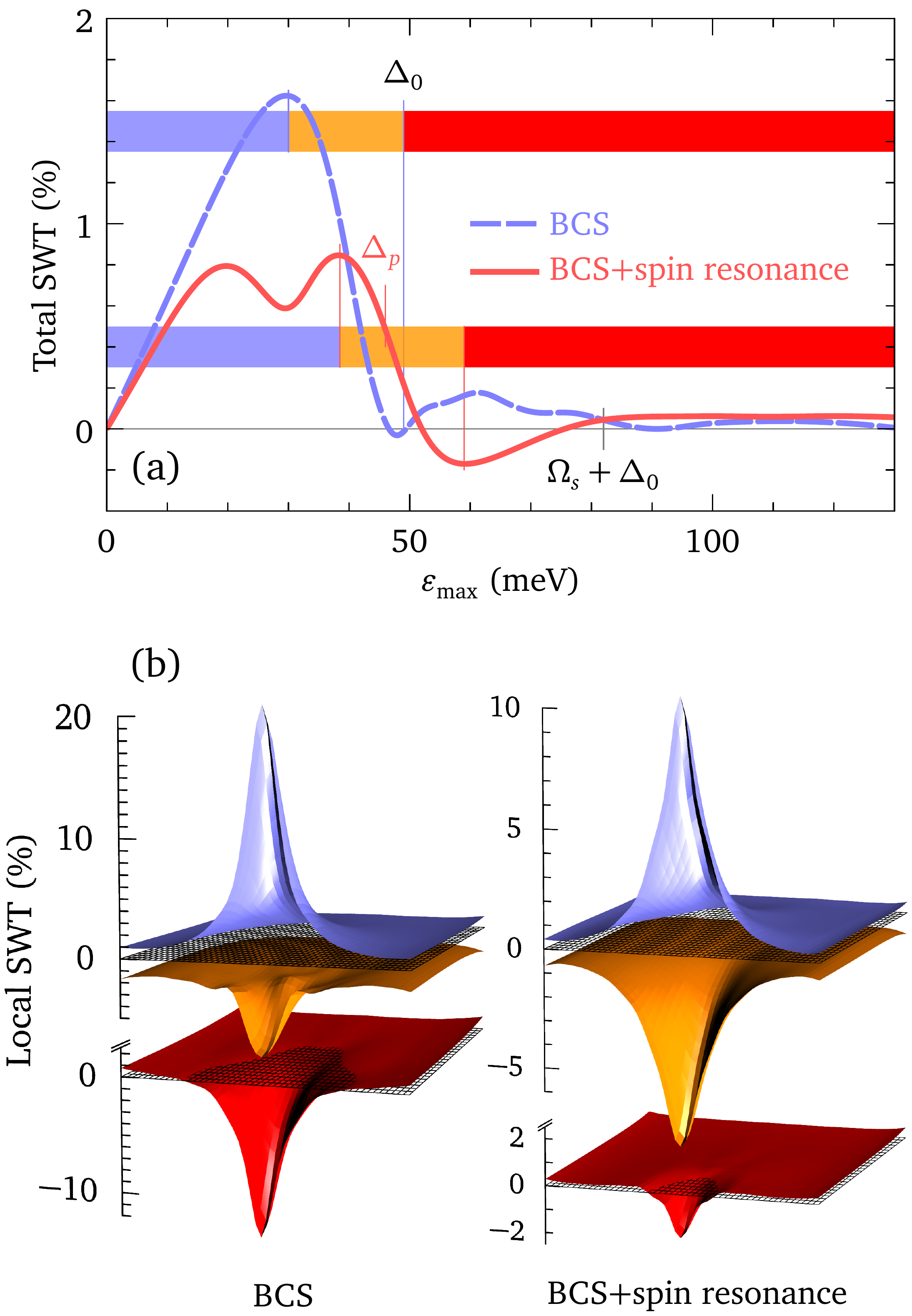}
\caption{\label{fig:fig-weight-2}
(a) Total spectral-weight transfer induced by the vortex in the energy range $|\varepsilon|<\varepsilon_{\mathrm{max}}$, with and without coupling to the spin resonance. Three energy domains are defined in each case: below the maximum transfer (light-blue), between the maximum and the minimum (orange) and the high-energy region (red). The local spectral-weight transfers for the three domains are shown in (b).
}
\end{figure}

\begin{figure*}[tb]
\includegraphics[width=1.8\columnwidth]{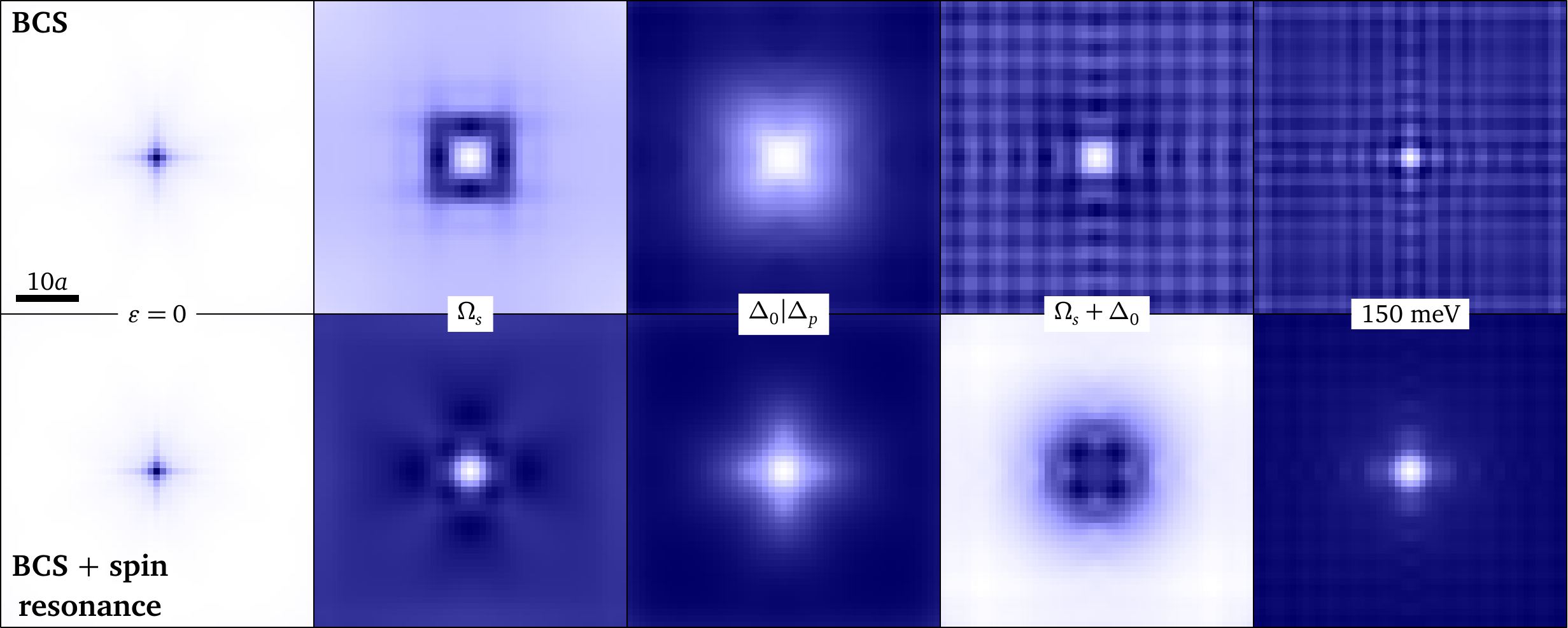}
\caption{\label{fig:fig-maps}
LDOS $N(\vec{r},\varepsilon)$ at several characteristic energies in the presence (bottom) and absence (top) of coupling to the spin resonance. The color scale goes from white (low) to dark blue (high) and covers the whole data range in each panel, such that absolute intensities in different panels cannot be compared. \cite{Note6}
}
\end{figure*}

The LDOS curves calculated in the region of the core with and without the coupling to spin fluctuations are compared in Figs.~\ref{fig:fig-core}--\ref{fig:fig-maps}. Figure~\ref{fig:fig-core} emphasizes the spectral differences at the vortex center, Fig.~\ref{fig:fig-traces} compares two spectral traces along the directions $(1,0)$ and $(1,1)$, Figs.~\ref{fig:fig-weight-1} and \ref{fig:fig-weight-2} illustrate the transfer of spectral weight, and Fig.~\ref{fig:fig-maps} compares LDOS spatial maps at fixed energies.

The interaction with the spin resonance leads to additional structure in the spectra. The rearrangement of spectral weight can be qualitatively understood along the lines given in Ref.~\onlinecite{Levy-2008}: additional dips not present in the BCS spectrum correspond to energies where the scattering rate is large, due to enhanced emission or absorption of spin fluctuations. The additional \emph{peaks} are less informative, because the spectral weight they carry is the one expelled from the dips towards both higher and lower energies. Schematically, the emission is strong at energies $\varepsilon>\Omega_s$ if the BCS DOS at energy $\varepsilon-\Omega_s$ is large and, inversely, the absorption is strong at $\varepsilon<-\Omega_s$ if the BCS DOS at $\varepsilon+\Omega_s$ is large. The strong emission region marked as \textbf{e$_1$} in Fig.~\ref{fig:fig-core} and characterized by a significant removal of spectral weight between $\sim20$ and 60~meV, as well as the absorption region marked \textbf{a$_1$}, result from the zero-bias peak in the BCS spectrum: quasiparticles at these energies decay into the near-zero energy vortex-core states. Similarly, the absorption and emission dips at \textbf{a$_2$} and \textbf{e$_2$} correspond to quasiparticles decaying into states at the gap edges, which survive as weak peaks at energies slightly lower than $\Delta_0$ in the BCS vortex-core spectrum. The peaks between \textbf{a$_1$} and \textbf{a$_2$} and between \textbf{e$_1$} and \textbf{e$_2$} collect part of the spectral weight removed from the corresponding dips, but it is also seen that a great part of this weight is transferred to energies larger than $\pm100$~meV. The small peaks right at \textbf{a$_1$} and \textbf{e$_1$} seem more difficult to assess, but an inspection of the spectral traces in Fig.~\ref{fig:fig-traces} reveals that they mark the onset of scattering at $\pm\Omega_s$. Below this threshold, the imaginary part of the self-energy (\ref{eq:Sigma2}) vanishes (see Fig.~\ref{fig:fig-Sigma}) and consequently the levels are renormalized to lower energy, but not broadened. In zero field or outside the vortex, the scattering rate tracks the linear increase of the $d$-wave BCS DOS and grows roughly linearly for $|\omega|>\Omega_s$, the Kramers-Kronig related renormalization bends over with a weaker slope and as a result the coherence peaks develop a shoulder (white arrows in Fig.~\ref{fig:fig-traces}). In the vortex core, however, the scattering rate jumps at $|\omega|=\Omega_s$ due to the zero-bias peak in the BCS spectrum and decreases for $|\omega|>\Omega_s$ (Fig.~\ref{fig:fig-Sigma}); the renormalization drops abruptly and changes sign, explaining the peaks at \textbf{a$_1$} and \textbf{e$_1$}.

The spectral traces shown in Fig.~\ref{fig:fig-traces} confirm the trends observed at the vortex center. The LDOS has more structure in the presence than in the absence of the coupling and dips in the interacting LDOS can be traced back to peaks in the BCS LDOS. On leaving the core, the dips at \textbf{a$_2$} and \textbf{e$_2$} become the usual dips of the zero-field spectrum (inset of Fig.~\ref{fig:fig-core}). A tiny shift of these dips to lower binding energy in the core follows the tiny shift of the BCS coherence peaks. The coherence peaks themselves, which are washed out in the core due to both loss of superconducting coherence and increased scattering, start to develop near $\Delta_p=46$~meV for $r>3a$, where the modulus of the pair potential is close to its asymptotic value (Fig.~\ref{fig:fig-gap}). The small peaks at \textbf{a$_1$} and \textbf{e$_1$} become the shoulders on the coherence peaks as already discussed. Finally, the zero-bias peak, depleted form much of its spectral weight, splits into several structures reminiscent of the Caroli--de Gennes--Matricon bound states, which disperse with increasing distance from the center and whose intensity decreases over a length larger than the core size. Using the Chebyshev-expansion method on a large $1001\times1001$ cluster, we have found that the wiggles in the BCS spectrum at energies above $\Delta_0$ in the core are finite-size artifacts: the converged spectrum is smooth at these energies. However, all little peaks at sub-gap energies in the BCS spectra of Fig.~\ref{fig:fig-traces} are real. The energies and the amplitudes of these peaks depend on the band-structure parameters. A similar verification is not possible with the coupling turned on, but we believe that the sub-gap peaks in the interacting LDOS are real spectral features as well.

It is interesting to study the spatial and energy dependence of the spectral-weight redistribution. To this end, we compare the spectral-weight transfer (SWT) induced by the spin resonance in zero field and around the vortex. In zero field, the SWT is defined in an energy domain $\mathscr{E}$ as \footnote{In the discussion of the spectral-weight transfer, we consider symmetric energy windows, thus ignoring subtleties associated with the breaking of particle-hole symmetry.}
	\begin{equation*}
		\Delta W=
		\frac{1}{W}\int_{|\varepsilon|\in\mathscr{E}}\hspace{-2mm}d\varepsilon\,N(\varepsilon)-
		\frac{1}{W_{\mathrm{BCS}}}\int_{|\varepsilon|\in\mathscr{E}}\hspace{-2mm}
		d\varepsilon\,N_{\mathrm{BCS}}(\varepsilon),
	\end{equation*}
where $W$ and $W_{\mathrm{BCS}}$ are the total spectral weights, defined for our purposes as $W=\int_{|\varepsilon|<200~\mathrm{meV}}d\varepsilon\,N(\varepsilon)$. The variation of $\Delta W$ as a function of $\varepsilon_{\mathrm{max}}$ for $\mathscr{E}=[0,\varepsilon_{\mathrm{max}}]$ is displayed in Fig.~\ref{fig:fig-weight-1}(a) as the black line. Not much happens at sub-gap energies. The weight removed at the lowest energies is overcompensated, such that some weight is gained at $\Omega_s$ and slightly above: this is the shoulder on the coherence peaks. The action really starts above $\Delta_p$, where 10\% of the weight is removed over 50~meV and recovered at energies higher than 100~meV: this is the dip. In the vortex, we repeat the analysis by replacing the zero-field DOS by the average vortex LDOS over a square extending from $-L$ to $+L$ in both directions. There is a range of $L$ values, between $20a$ and $30a$, where the result is almost independent of $L$ [Fig.~\ref{fig:fig-weight-1}(a)]. We therefore set $L=25a$ in the following. Smaller values of $L$ focus too much on the core region, much larger values, if accessible, would mask the signature of the vortex and approach the zero-field result. With $L=25a$, there is compensation of the average SWT very close to the two characteristic energies $\Omega_s$ and $\Delta_p$. The question arises, whether some weight is redistributed \emph{spatially} among neighboring sites: the answer is yes. In Fig.~\ref{fig:fig-weight-1}(b), the local SWT is plotted in four energy domains. The local SWT is defined as 
	\begin{equation*}
		\Delta W(\vec{r})=
		\frac{1}{\bar{W}}\int_{|\varepsilon|\in\mathscr{E}}\hspace{-2mm}d\varepsilon\,N(\vec{r},\varepsilon)-
		\frac{1}{\bar{W}_{\mathrm{BCS}}}\int_{|\varepsilon|\in\mathscr{E}}\hspace{-2mm}
		d\varepsilon\,N_{\mathrm{BCS}}(\vec{r},\varepsilon),
	\end{equation*}
with $\bar{W}$ and $\bar{W}_{\mathrm{BCS}}$ the average integrated spectral weights in the spatial region considered ($L=25a$). At low energy $|\varepsilon|<\Omega_s$, a SWT of the order of 5\% occurs from the vortex center to the surroundings. The depleted region is more extended than the core where the pair-potential modulus is significantly suppressed. The loss of low-energy weight in the vortex is obvious in Figs.~\ref{fig:fig-core} and \ref{fig:fig-traces}, but the spatial compensation which occurs at distances $r\gtrsim15a$ cannot be seen on those figures. Between $\Omega_s$ and $\Delta_p$, there is again a net SWT from inside the vortex to outside, although much weaker. The two small maxima at the vortex center in Fig.~\ref{fig:fig-weight-1}(b) are associated with the gain of weight near \textbf{a$_1$} and \textbf{e$_1$} (see Fig.~\ref{fig:fig-core}). Therefore, although there is no global SWT at the energy $\Delta_p$, locally there is a deficiency of weight in the vortex core and an excess outside. This unbalance is compensated in the region of the dip: between $\Delta_p$ and 100~meV, there is a loss of spectral weight everywhere, but much less in the core than outside, such that at 100~meV, the global loss of weight is nearly uniform in space, as confirmed by the nearly uniform recovery above 100~meV.

\footnotetext{See ancillary file at \url{http://arxiv.org/abs/1508.04225} for an animated version of Fig.~\ref{fig:fig-maps} showing data between $-200$ and $+200$~meV with relative and absolute color scales.}

From an experimental perspective, the SWT induced by the spin resonance is not accessible and thus not a quantity of particular interest: its measurement would require to turn off the interaction with the spin resonance. However, the SWT induced by the vortex itself is, in principle, directly accessible with nowadays technology, by switching the magnetic field on and off and measuring the LDOS with and without the vortex. The local vortex-induced SWT is defined as
	\begin{equation*}
		\Delta W(\vec{r})=
		\frac{1}{\bar{W}}\int_{|\varepsilon|\in\mathscr{E}}\hspace{-2mm}d\varepsilon\,N(\vec{r},\varepsilon)-
		\frac{1}{W}\int_{|\varepsilon|\in\mathscr{E}}\hspace{-2mm}d\varepsilon\,N_0(\varepsilon).
	\end{equation*}
with $N(\vec{r},\varepsilon)$ the LDOS of the vortex and $N_0(\varepsilon)$ the zero-field DOS. The total vortex-induced SWT is obtained by replacing $N(\vec{r},\varepsilon)$ by its average in the same spatial region as above ($L=25a$). The result is displayed in Fig.~\ref{fig:fig-weight-2} for the cases with and without the coupling to the spin resonance. There are qualitative differences, offering a chance to distinguish experimentally the two situations. The total SWT shows an accumulation of states at low energy [Fig.~\ref{fig:fig-weight-2}(a)]. Part of these states are outside the vortex core [Fig.~\ref{fig:fig-weight-2}(b)] and correspond to the increased DOS due to the Doppler shift of the dispersion \cite{Volovik-1993}; part of them correspond to the zero-bias peak in the core. Both the Doppler-shift and the core contributions are reduced by a factor close to two when the interaction in present. This is consistent with a renormalization of the quasiparticle velocity by a factor $1/(1+\lambda)$, since the renormalization factor $\lambda$ is close to unity \cite{Berthod-2013}. The Doppler-shift approximation is valid at low energy and indeed one sees that it breaks down when approaching the gap scale. In this energy domain, the LDOS outside the vortex is \emph{lower} than in the absence of field and as a result the total SWT decreases. In the BCS case, the compensation is complete at the energy $\Delta_0$. This is only true on average, though, because locally there is still excess weight in the core and deficiency outside at the energy $\Delta_0$. This is restored by a very nonuniform SWT at higher energy, as seen in Fig.~\ref{fig:fig-weight-2}(b). Here we observe the most significant differences. With the coupling on, there is no compensation on average at the peak energy $\Delta_p$ and the local compensation occurring near 60~meV at the minimum of the total SWT is almost complete, such that the high-energy SWT is small and much more uniform spatially than in the BCS case.

\begin{figure*}[tb]
\includegraphics[width=1.5\columnwidth]{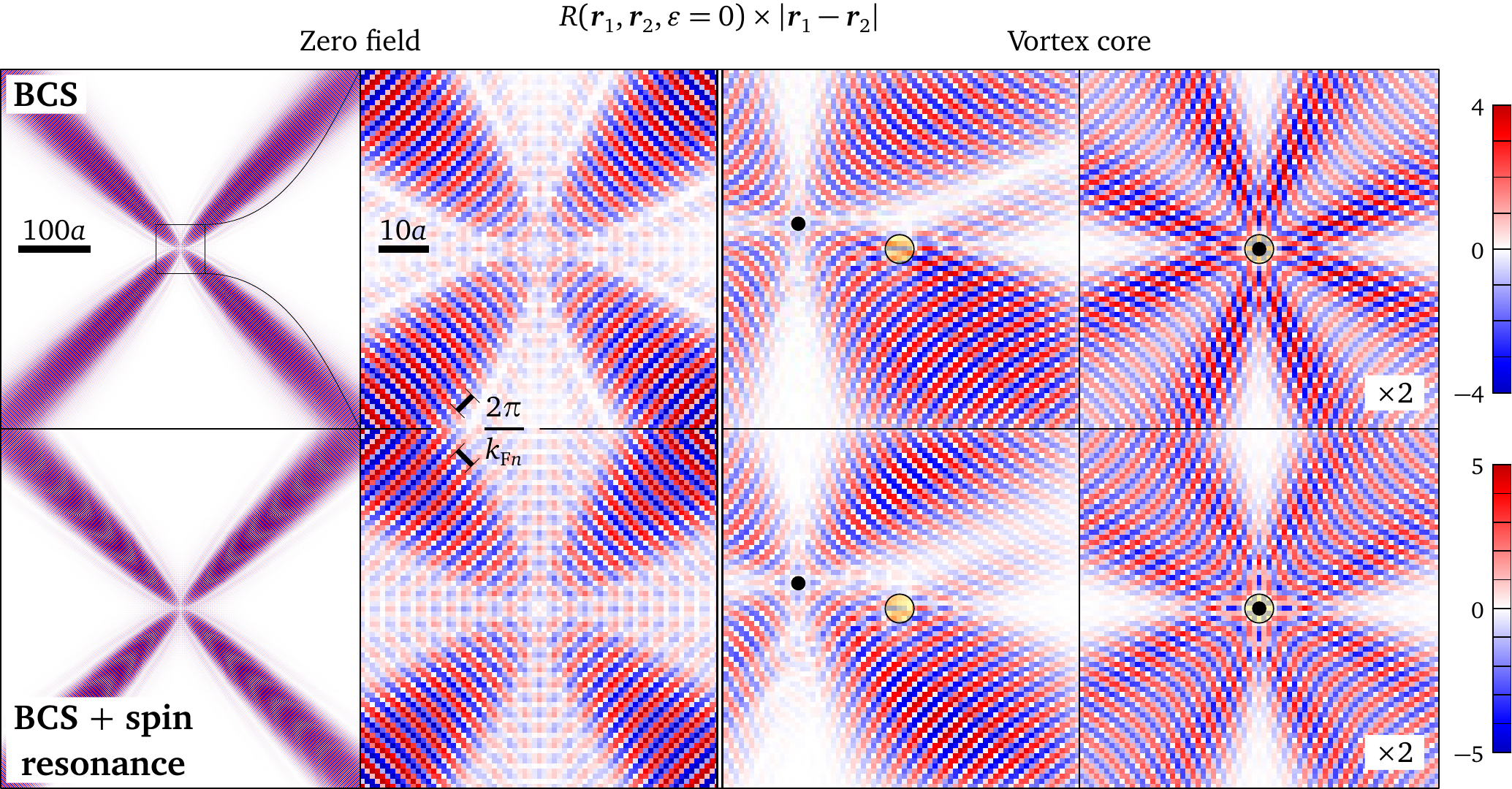}
\caption{\label{fig:fig-nonlocal}
Nonlocal conductance ratio at zero energy in zero field (columns 1 and 2) and close to a vortex (columns 3 and 4). $\vec{r}_1$ is fixed to $(0,0)$, corresponding to the center of the panel, in columns 1, 2, and 4; $\vec{r}_1=(-20,5)a$ in column 3 (black dot). The yellow circles indicate the vortex center. All panels in a row share the same color scale displayed on the right.
}
\end{figure*}

We close this section by comparing in Fig.~\ref{fig:fig-maps} the spatial LDOS distributions in the presence and in the absence of the coupling, at a few characteristic positive energies. Similar behaviors are found at negative energies. The most significant differences occur at $\varepsilon=\Omega_s$ and $\varepsilon=\Omega_s+\Delta_0$. The zero-energy states are similar in both cases: they are localized on a few lattice sites and spread along the principal axes, not along the nodal directions. At $\varepsilon=\Omega_s$, the BCS case shows a square, which resembles a Caroli--de Gennes--Matricon state. The radius of such a state at energy $E$ may be estimated as \cite{Berthod-2005} $r_E=(E/\Delta_0)[1+2r_c/(\pi\xi)]r_c$, where $r_c$ represents the core radius. For $E=\Omega_s$, this expression gives the value $r_{\Omega_s}\sim5a$ indicated by the data if one takes $r_c=3.9a$, which is a reasonable value, considering the gap profile shown in Fig.~\ref{fig:fig-gap}. The spatial structure of the LDOS is completely changed by the spin resonance: the density is spread out of the vortex (see also Fig.~\ref{fig:fig-weight-1}), such that there is no (quasi-)localized state at this energy. At the gap edge, the LDOS shows in both cases a depression in the core, but the latter appears to be shrunk and rotated by 45$^\circ$ by the spin resonance. The difference is most spectacular at the dip energy $\Omega_s+\Delta_0$. In the BCS case, there is a deficiency of density in the core (see also Fig.~\ref{fig:fig-weight-2}) and the LDOS displays quasiparticle interference patterns. The latter are due to scattering on the vortex, but also on the cluster's boundaries. We have checked this by repeating the calculation in a much bigger cluster with the Chebyshev-expansion method. These interference patterns disperse with energy, as seen in the 150~meV map. They are also present a lower energies, but too weak to be clearly resolved in comparison with the vortex-related structures. With the spin resonance, there appears to be an excess rather than a deficiency of density in the core. As already discussed, $\Omega_s+\Delta_0$ is the energy where the scattering rate is largest outside the vortex, due to the BCS coherence peaks at $\Delta_0$. Although also large, the scattering rate is about two times smaller in the core where these coherence peaks are reduced (see Fig.~\ref{fig:fig-Sigma}). Hence less density is removed in the core than outside. It should be noted that this difference is small, of the order of 6\%, but is magnified by the choice of the color scale in Fig.~\ref{fig:fig-maps}. For comparison, the LDOS difference at $\varepsilon=0$ between the maximum and the minimum is $\sim90$\% \cite{Note6}. Even more striking at $\Omega_s+\Delta_0$ is the complete washing out of the interference patterns by the spin resonance. This suggests that the mean free path does not exceed the quasiparticle wavelength. We may tentatively estimate the mean free path as $\ell=\langle v_{\mathrm{F}}^*\rangle\tau$, where $\langle v_{\mathrm{F}}^*\rangle$ is the average renormalized Fermi velocity, which is $1.15\times10^7$~cm/s, $\tau=\hbar/(2Z\gamma)$ is the quasiparticle lifetime, $\gamma$ being the scattering rate, and $Z=\langle v_{\mathrm{F}}^*\rangle/\langle v_{\mathrm{F}}\rangle$ the quasiparticle residue. Outside the vortex, $\gamma$ is typically $100$~meV at the energy $\Omega_s+\Delta_0$ and 50~meV at higher energies (Fig.~\ref{fig:fig-Sigma}). The corresponding values of $\ell$ are $2.3a$ and $4.5a$, respectively. The former is shorter than the period of the BCS quasiparticle oscillations in Fig.~\ref{fig:fig-maps}, while the latter is comparable. This may explain why no interference at all is seen at $\Omega_s+\Delta_0$, while some traces remain at 150~meV. Our naive formula underestimates the mean free path, however: a more rigorous evaluation, to be performed in the next section, gives values nearly five times larger.

\subsection{Bogoliubov quasiparticles in real space}\label{sec:NLCR}

The quasiparticle LDOS discussed in the previous section can be measured using a scanning tunneling microscope (STM) \cite{Fischer-2007}. There are other important quasiparticle properties which are not visible in the LDOS alone. For instance, their nodal character: the fact that the zero-energy LDOS in Fig.~\ref{fig:fig-maps} extends along the lattice axes rather than along the diagonal does not imply that the zero-energy quasiparticles in the vortex have lost their nodal character. This nodal character is still present, as we will see. As another example, the short mean free path of the quasiparticles coupled to the spin resonance is not directly apparent in the LDOS, but only indirectly through its effect on the interference patterns. It is therefore useful to go beyond the LDOS and to consider nonlocal effects that reveal directly these quasiparticle properties. While the LDOS is encoded in the diagonal elements $G(\vec{r},\vec{r},\varepsilon)$ of the retarded Green's function $G$, the off-diagonal elements $G(\vec{r},\vec{r}',\varepsilon)$ contain all additional information about the single-particle excitations. It is well established that these off-diagonal terms are accessible experimentally, for instance by local double-tip tunneling \cite{Byers-1995, Niu-1995}. A double-tip STM with nanometer distances between the tips has already been demonstrated \cite{Xu-2006}. Here, we consider the quantity
	\begin{equation}\label{eq:conductance-ratio}
		R(\vec{r}_1,\vec{r}_2,\varepsilon)=\frac
		{\mathrm{Im}[G(\vec{r}_1,\vec{r}_2,\varepsilon)+G(\vec{r}_2,\vec{r}_1,\varepsilon)]}
		{\mathrm{Im}[G(\vec{r}_1,\vec{r}_1,\varepsilon)+G(\vec{r}_2,\vec{r}_2,\varepsilon)]}.
	\end{equation}
We show in Appendix~\ref{app:double-tip} that this quantity is the relative change of the tunneling conductance measured at zero temperature with two coupled tips at points $\vec{r}_1$ and $\vec{r}_2$, with respect to the conductance measured with two uncoupled tips at the same two positions. We refer to this as the nonlocal conductance ratio (NLCR). In the proposed setup, the two tips are coherently connected to the same reservoir. As a result, the NLCR can be positive or negative, unlike in the setup of Ref.~\onlinecite{Niu-1995}, where the ``transconductance'' is positive definite. In a translation-invariant system, $R(\vec{r}_1-\vec{r}_2,\varepsilon)$ is proportional to the imaginary part of the Fourier transform of the Green's function $G(\vec{k},\varepsilon)$. Because $G(\vec{k},\varepsilon)=G(-\vec{k},\varepsilon)$, this is identical to the Fourier transform of the imaginary part of $G(\vec{k},\varepsilon)$, i.e., the Fourier transform of the spectral function $A(\vec{k},\varepsilon)$. For $\varepsilon<0$, this quantity could in principle be deduced from photoemission data, but this has not been reported so far, to the best of our knowledge.

\begin{figure*}[tb]
\includegraphics[width=1.5\columnwidth]{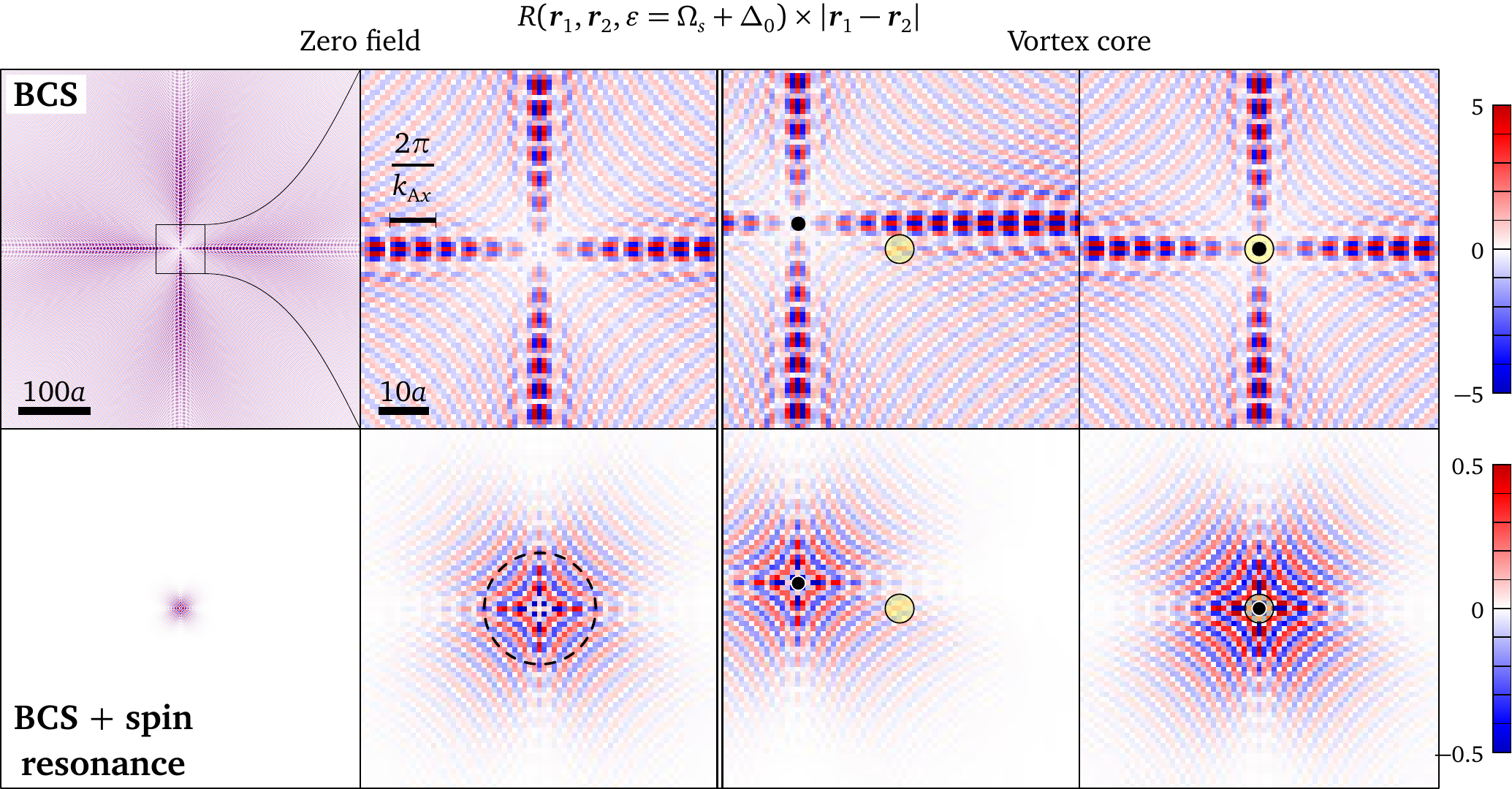}
\caption{\label{fig:fig-nonlocal2}
Same as Fig.~\ref{fig:fig-nonlocal}, at the energy $\varepsilon=\Omega_s+\Delta_0$. The dashed circle indicates the mean free path, calculated by fitting an exponential decay to the NLCR.
}
\end{figure*}

The NLCR at zero energy, calculated without magnetic field as well as close to a vortex, are compared in Fig.~\ref{fig:fig-nonlocal}. Since $R(\vec{r}_1,\vec{r}_2,\varepsilon)$ has a geometrical $1/r$ decay for long distances, we multiply by $|\vec{r}_1-\vec{r}_2|$ for plotting. In zero field, the NLCR is shown for large distances (first panel in each row) and for short distances over the area corresponding to the cluster size of the vortex calculation (second panel in each row). The nodal character and the long-range nature of the zero-energy excitations appear very clearly in this representation, both without and with the coupling to the spin resonance. The period of oscillation corresponds to the nodal Fermi wavelength, as indicated in the figure. In the BCS case, the Fermi wavelength is almost exactly commensurate with the lattice, $2\pi/\vec{k}_{\mathrm{F}n}=3.003(a,a)$. It is renormalized to slightly lower values $2\pi/\vec{k}_{\mathrm{F}n}^*=2.961(a,a)$ by the spin resonance: this small incommensurability explains the Moir{\'e} pattern with a period $\sim36a$ along the diagonals.

In order to visualize excitations in the vicinity of the vortex core, we first take $\vec{r}_1=(-20,5)a$ and plot $R(\vec{r}_1,\vec{r}_2,0)$ as a function of $\vec{r}_2$. This may be loosely understood as the long-time behavior of the wave function for a particle created at $\vec{r}_1$. It is seen that the latter is similar to the zero-field quasiparticle. The spreading away from the nodal directions appears to be wider than in the zero-field case, but this is a finite-size effect. The scattering on the vortex is clearly visible; one also sees that it is a relatively weak effect. Note that the Doppler effect, by which all momenta are shifted by a quantity $(\hbar/2r)\vec{e}_{\vartheta}$, where $r$ is the distance to the vortex and $\vec{e}_{\vartheta}$ is the direction of the supercurrent, is too small to be seen. For $r\sim 20a$, the relative change of wavelength is $1/(1+2k_{\mathrm{F}}r)\sim 2\%$. When the particle is created right at the vortex center (rightmost panel in each row), the NLCR is globally smaller because the LDOS at this point is large. The excitation remains nevertheless nodal, despite some loss of nodal intensity (which is less pronounced in the presence of the spin resonance), in contrast to the LDOS, which extends along the antinodal directions at $\varepsilon=0$ (Fig.~\ref{fig:fig-maps}).

Figure~\ref{fig:fig-nonlocal2} shows the NLCR calculated at the energy $\varepsilon=\Omega_s+\Delta_0$, where the effects of the spin resonance are strongest. In the BCS case, the spectral weight is mostly in the antinodal region, such that the NLCR extends mainly along $(1,0)$ and $(0,1)$. At this energy, the antinodal wave-vector is $\vec{k}_{\mathrm{A}}=(0.22,1)\pi/a$, corresponding to a period of $9.1a$ along one direction and $2a$ along the other, as indicated in the figure. Some nodal states are also mixed in, but their contribution is weak. Note that the scattering on the vortex is inexistent at this energy, both in the BCS and spin-resonance cases. Three major differences are seen with the coupling to the spin resonance. First, the amplitude of nonlocal effects is reduced by an order of magnitude (see color scales on the right of Fig.~\ref{fig:fig-nonlocal2}). Second, the $(\pi,\pi)$ scattering suppresses spectral weight and broadens states in the antinodal region, such that the better-defined nodal states are dominating the shape of the NLCR. Third, the appearance of a mean free path is manifested by an exponential decay of the NLCR with increasing distance $|\vec{r}_1-\vec{r}_2|$. Fitting this exponential decay, we obtain a mean free path $\ell=11a$ (dashed circle in the figure), which is much larger than our previous estimate based on the average Fermi velocity and scattering rate. Repeating the same fitting as a function of energy, we find that the mean free path tracks the energy dependence of the average scattering rate, but with an exponent different from $-1$. Both quantities are approximately related by $\ell/a\approx 320\times(\gamma/\mathrm{meV})^{-2/3}$. In this analysis, the average scattering rate was defined as $\gamma(\varepsilon)=(1/2)\langle p(\vec{k},\varepsilon)\mathrm{Im}[-\Sigma_{11}(\vec{k},\varepsilon)-\Sigma_{22}(\vec{k},\varepsilon)]\rangle_{\mathrm{BZ}}$, with $p(\vec{k},\varepsilon)=A_0(\vec{k},\varepsilon)/\langle A_0(\vec{k},\varepsilon)\rangle_{\mathrm{BZ}}$, $A_0(\vec{k},\varepsilon)$ being the spectral function without pairing gap ($\Delta_{\vec{k}}=0$) and $\langle\cdots\rangle_{\mathrm{BZ}}$ standing for a Brillouin-zone average.

\section{Discussion}\label{sec:discussion}

The model (\ref{eq:Sigma1}) reproduces the zero-field STM data of Bi-2223 very well \cite{Berthod-2013}; yet, the same model (\ref{eq:Sigma2}) fails to reproduce the experimental vortex-core spectrum of this material \cite{Note1}. The latter is similar to published results for Bi-2212 \cite{Renner-1998b, Hoogenboom-2000a, Pan-2000b, Matsuba-2003a, *Matsuba-2007, Levy-2005, Yoshizawa-2013}, lacking a zero-bias peak in the core, but showing weak low-energy structures on top of a (pseudo-)gapped spectrum. The self-energy (\ref{eq:Sigma2}) does not suppress the zero-bias peak, although it reduces its weight; neither does it smear out the BCS spectra in a way which would turn the vortex-core spectrum into the structureless signature observed in dirty superconductors, but rather it adds structure to it. Lastly, while the experiment suggests that the dip is disappearing in the vortex core, it is not the case with the model (\ref{eq:Sigma2}). The fact that the model (\ref{eq:Sigma2}) yields a peak in the vortex core illustrates the robustness of this zero-bias feature, associated with the topological defect carried by the vortex. It is unlikely that any theory in which the zero-field spectrum is entirely made of (possibly damped) Bogoliubov quasiparticles of a BCS superconductor can produce a vortex-core LDOS without zero-bias peak, unless something that goes beyond the BCS theory happens in the vortices. The nucleation of a static antiferromagnetic order is one possibility \cite{Arovas-1997, Zhu-2001a, Takigawa-2003, *Takigawa-2004}, but a detailed comparison of this theory with STM data of the cuprates is still missing. Charge order in the vortices has also been suggested experimentally \cite{Wu-2013}, but the local spectral signatures of such an order remain so far unknown. While these interpretations focus on the possible non-BCS nature of the vortices, many experiments point to a non-BCS nature of the zero-field spectrum as well, in relation to the phenomenon of the pseudogap. In this perspective, the success of the model (\ref{eq:Sigma1}) in reproducing the zero-field data must be an accident, indicating that the general BCS--Eliashberg structure of the Green's function would be correct, despite the fact that the underlying physics would be wrong. The failure of (\ref{eq:Sigma2}) in the vortex core would then reveal the trick, because \emph{there} the BCS--Eliashberg structure, which assumes superconducting coherence of all excitations, would be inappropriate. Discriminating between the two scenario requires a complete microscopic theory of the pseudogap and its interaction with superconductivity, which is not yet available.

In ferropnictides, the role of antiferromagnetic spin fluctuations as the driving force for pairing is more firmly established than in the cuprates \cite{Inosov-2015}. Tunneling spectra in zero field and in vortices are available for a number of these materials and spectral structures suggestive of a coupling to the spin resonance, as measured by neutron scattering, are routinely observed \cite{Song-2011, Shan-2011, Shan-2012, Hanaguri-2012, Wang-2013, Note7}. \footnotetext{Note that the signature in the superconducting DOS of the coupling to a bosonic mode is a dip if the order parameter has $d_{x^2-y^2}$ symmetry, but a break (peak in the DOS derivative) if it has $s$ symmetry; see Ref.~\onlinecite{Berthod-2010}.} For example in Ba$_{0.6}$K$_{0.4}$Fe$_2$As$_2$, the $(\pi,\pi)$ resonance is measured at 14~meV \cite{Christianson-2008}, the tunneling spectrum shows a gap of 7~meV and presents the expected DOS structure near 21~meV \cite{Wang-2013}. The evidence of the coupling in the vortex-core spectra is more elusive, but encouraging. In FeSe, for instance, the vortex-core spectrum reported in Fig.~2A of Ref.~\onlinecite{Song-2011} has a lot in common with that displayed in Fig.~\ref{fig:fig-core}. The weak structures at $\pm 3$~meV in the former are reminiscent of the ones observed at the onset of scattering in the latter (\textbf{a$_1$} and \textbf{e$_1$}); the structures at $\pm5$~meV correspond to the main absorption and emission at \textbf{a$_2$} and \textbf{e$_2$}. These numbers are consistent with the value of the zero-field gap, observed near 2~meV, and the value of the spin resonance energy which, although not yet measured by neutron scattering, is expected near $4.4k_{\mathrm{B}}T_c=3$~meV in this material \cite{Wang-2013, Inosov-2015}. Further evidence could emerge from a careful study of the spatial structure of the LDOS, or from an analysis of the vortex-induced spectral weight transfer.

The study of quasiparticle interference (QI) by STM has been a fruitful development in the last decade. QI does not image the quasiparticles directly, but their interferences due to multiple scattering on defects: this is an advantage, because the interferences are long-ranged even if the quasiparticles are not, but it is also a weakness of the method, which makes it impractical for clean systems. Measuring the nonlocal conductance ratio (NLCR) introduced in Sec.~\ref{sec:NLCR} would provide additional information about the quasiparticles. This is a considerable challenge by tunneling, perhaps less so by photoemission, although with the latter technique the local information is lost and positive energies are not accessible. The NLCR exists also in clean systems and images the quasiparticles directly, showing their momenta---while QI shows differences of momenta---as well as their spatial coherence range. Via the NLCR, local tunneling can probe the reciprocal space with two advantages compared to photoemission: an easy access to positive energies and the possibility to follow local variations of the reciprocal-space properties.

\vspace{-2em}
\section{Conclusions}

We have calculated the electronic properties of a vortex in a two-dimensional one-band $d$-wave superconductor, in which the Bogoliubov quasiparticles interact with a spin resonance centered at momentum $(\pi,\pi)$. Unlike previous calculations dealing with a static antiferromagnetic order in the vortex core, the dynamical case considered here is not amenable to a mean-field treatment, but requires to evaluate a self-energy that is nonlocal in space and time. We have discussed several signatures of the coupling to the spin resonance in the local density of states near the core. In spite of the fact that the model we use fits quantitatively the tunneling spectrum of Bi$_2$Sr$_2$Ca$_2$Cu$_3$O$_{10+\delta}$ in zero field, it fails to reproduce the peculiar vortex-core spectra of the cuprates. We believe that passing the test of the vortex-core spectrum is a tough sanity check for all theories of the cuprates electronic structure. Our results may nevertheless be useful to understand the vortex-core tunneling spectra of iron-based superconductors and we have argued that some of the signatures we discuss may be present in published data for FeSe.

Bogoliubov quasiparticles coupled to spin fluctuations loose spatial coherence. We have shown that this is manifested by an extinction of quasiparticle interference at the energies where the effect of the coupling is strongest. We have also discussed a new way to look at quasiparticles in real space, which allows one to access quasiparticle properties that are not directly visible in the LDOS, such as their nodal/antinodal character and their mean free path.

\acknowledgments

This work was supported by the Swiss National Science Foundation under Division II. The calculations were performed in the University of Geneva with the clusters Mafalda and Baobab.

\vspace{-1em}
\appendix

\section{Symmetry properties of the Green's function}
\label{app:symmetry}

In the absence of coupling to spin fluctuations ($\Sigma_{ij}\equiv0$), the Nambu-Gorkov equation (\ref{eq:Gorkov}) reads
	\[
		\begin{pmatrix}\mathscr{G}_0^{-1}(z)&-\Delta\\ -\Delta^{\dagger}&
		-[\mathscr{G}_0^{-1}(-z)]^T\end{pmatrix}\begin{pmatrix}\mathscr{G}_{11}(z)&
		\mathscr{G}_{12}(z)\\ \mathscr{G}_{21}(z)&\mathscr{G}_{22}(z)\end{pmatrix}=
		\begin{pmatrix}\openone&0\\0&\openone\end{pmatrix},
	\]
with the explicit solution
	\begin{align*}
		\mathscr{G}_{11}(z)
			&=\{\mathscr{G}_0^{-1}(z)+\Delta\mathscr{G}_0^T(-z)\Delta^{\dagger}\}^{-1}\\
			\mathscr{G}_{12}(z)
			&=-\{[\mathscr{G}_0(z)\Delta\mathscr{G}_0^T(-z)]^{-1}+\Delta^{\dagger}\}^{-1}\\
			\mathscr{G}_{21}(z)
			&=-\{[\mathscr{G}_0^T(-z)\Delta^{\dagger}\mathscr{G}_0(z)]^{-1}+\Delta\}^{-1}\\
			\mathscr{G}_{22}(z)
			&=-\{[\mathscr{G}_0^T(-z)]^{-1}+\Delta^{\dagger}\mathscr{G}_0(z)\Delta\}^{-1}.
		\end{align*}
The property $\mathscr{G}_0(z)=\mathscr{G}_0^{\dagger}(z^*)$, easily checked from the Fourier representation of $\mathscr{G}_0$, implies that
	\[
		\mathscr{G}_{11}(z)=\mathscr{G}_{11}^{\dagger}(z^*),\quad
		\mathscr{G}_{12}(z)=\mathscr{G}_{21}^{\dagger}(z^*),\quad
		\mathscr{G}_{22}(z)=\mathscr{G}_{22}^{\dagger}(z^*).
	\]
Furthermore, the property $\Delta=\Delta^{\!T}$, which follows from the symmetry of the pairing interaction, allows one to deduce two additional relations:
	\[
		\mathscr{G}_{22}(z)=-\mathscr{G}_{11}^*(-z^*),\quad
		\mathscr{G}_{12}(z)=\mathscr{G}_{21}^*(-z^*).
	\]

\section{Chebyshev expansion of the Green's functions}
\label{app:Chebyshev}

Consider a one-band quadratic Hamiltonian defined on a lattice spanned by the discrete vectors $\vec{r}$. In the superconducting state, there are two degrees of freedom at each lattice site, namely the Bogoliubov--de Gennes amplitudes $[u(\vec{r}),v(\vec{r})]$. The real-space retarded Green's function at energy $E$ is $G(\vec{r},\vec{r}',E)=\langle\vec{r}|[(E+i0^+)\openone-H]^{-1}|\vec{r}'\rangle$. In this expression, $H$ is the matrix representing the Hamiltonian and the notation $|\vec{r}\rangle$ means a state describing an electron localized at point $\vec{r}$: the corresponding state vector has $u(\vec{r})=1$ and all other components equal to zero. For each pair $(\vec{r},\vec{r}')$, the Hamiltonian is the $2\times2$ block
	\begin{equation}\label{eq:block}\begin{pmatrix}
		-\mu\delta_{\vec{r}\vec{r}'}+t_{\vec{r}\vec{r}'} & \Delta(\vec{r},\vec{r}') \\[1em]
		\Delta^*(\vec{r}',\vec{r}) & \mu\delta_{\vec{r}\vec{r}'}-t^*_{\vec{r}\vec{r}'}
	\end{pmatrix}.\end{equation}
The expansion method of Ref.~\onlinecite{Covaci-2010} consists in expanding the matrix $G(E)=[(E+i0^+)\openone-H]^{-1}$ on Chebyshev polynomials in $H$. The Chebyshev polynomials are $T_n(x)=\cos(n\arccos x)$, defined in the range $-1\leqslant x\leqslant 1$. The expansion of $G(E)$ on these polynomials reads
	\begin{equation}\label{eq:expansion}
		G(E)=\frac{1}{a}\sum_{n=0}^{\infty}
		\frac{i(\delta_{n0}-2)e^{-in\arccos(\tilde{E})}}{\sqrt{1-\tilde{E}^2}}T_n(\tilde{H}).
	\end{equation}
Tildes indicate rescaled dimensionless energies, which fall within the range $[-1,1]$ on which the Chebyshev polynomials are defined: $\tilde{H}=(H-b)/a$, $\tilde{E}=(E-b)/a$, with $a$ and $b$ the width and center of the spectrum of $H$, respectively (exact values are not required). The polynomials obey the recursion relation $T_{n+1}(x)=2xT_n(x)-T_{n-1}(x)$, such that the evaluation of the matrix elements $\langle\vec{r}|T_n(\tilde{H})|\vec{r}'\rangle$ can be performed iteratively. This computational method is very appealing for several reasons. First, the calculation of the matrix elements only requires to evaluate matrix-vector products $H|\psi\rangle$, with no need to store the Hamiltonian in memory; only three state vectors need to be stored in the recursive scheme: $|\psi_0\rangle=|\vec{r}'\rangle$,  $|\psi_1\rangle=\tilde{H}|\vec{r}'\rangle$, $|\psi_{n+1}\rangle=2\tilde{H}|\psi_{n}\rangle-|\psi_{n-1}\rangle$. Second, when the matrix elements are known, the Green's function can be computed at any energy $E$ in almost no time. Third, the calculation is trivially parallel in the positions $\vec{r}$. Lastly, as the Hamiltonian propagates the initial state $|\vec{r}'\rangle$ on the neighboring sites of $\vec{r}'$ and so on at each iteration, the linear lattice size needed to calculate the matrix element $\langle\vec{r}|\tilde{H}^n|\vec{r}'\rangle$ is proportional to $n$. Nevertheless, manageable lattice sizes give accurate results thanks to the good convergence properties of the Chebyshev expansion.

When the $n$ sum in Eq.~(\ref{eq:expansion}) is truncated to some maximal value $N$, the linear lattice size $M$ must ideally be such that boundary effects do not affect the last matrix element $\langle\vec{r}|T_N(\tilde{H})|\vec{r}'\rangle$. For instance, if the propagation proceeds only through hopping to the nearest neighbors, it takes $N=2M$ iterations until the reflection from the boundary propagates back to the site $\vec{r}$ and $N=4M$ iterations until interferences between the reflections on the two opposite boundaries can be felt at $\vec{r}$ (assuming that $\vec{r}$ is the central site of the system, which can always be arranged). If the minimal size requirements are met, there remain nevertheless oscillations due to the truncation itself, known as Gibbs oscillations. Those can be suppressed by well-known procedures \cite{Weisse-2006}. We use the Lorentz kernel, which amounts to multiplying each term of the sum in Eq.~(\ref{eq:expansion}) by $\sinh(\tilde{\Gamma}N-\tilde{\Gamma}n)/\sinh(\tilde{\Gamma}N)$ and is equivalent to introducing a phenomenological Dynes scattering rate $\Gamma=a\tilde{\Gamma}$.

The BCS gap equation is $\Delta(\vec{r},\vec{r}')=-V(\vec{r},\vec{r}')\langle c_{\vec{r}\uparrow}c_{\vec{r}'\downarrow}\rangle$, where $V(\vec{r},\vec{r}')$ is the pairing interaction and $c_{\vec{r}\sigma}$ annihilates a spin-$\sigma$ electron at position $\vec{r}$. The average value $\langle c_{\vec{r}\uparrow}c_{\vec{r}'\downarrow}\rangle$ can be related to the retarded anomalous function $F^+(\vec{r},\vec{r}',E)=\langle\bar{\vec{r}}|G(E)|\vec{r}'\rangle$. $|\bar{\vec{r}}\rangle$ is the state vector describing one hole localized at point $\vec{r}$, i.e., with $v(\vec{r})=1$ and all other components equal to zero. We obtain
	\begin{multline}\label{eq:gap}
		\Delta(\vec{r},\vec{r}')=V(\vec{r},\vec{r}')
		\int_{-\infty}^{\infty} dE\,f(E)\\\times\frac{-i}{2\pi}\left[
		F^+(\vec{r},\vec{r}',E)-F^+(\vec{r}',\vec{r},-E)\right]^*,
	\end{multline}
where $f(E)=(e^{E/k_{\mathrm{B}}T}+1)^{-1}$ is the Fermi function. Setting the temperature to zero, inserting the Chebyshev expansion of $F^+$ into Eq.~(\ref{eq:gap}) and performing analytically the integral, we are led to the following expression:
	\begin{multline}\label{eq:gapequation}
		\Delta(\vec{r},\vec{r}')=V(\vec{r},\vec{r}')\frac{i}{\pi}\sum_{n=1}^{\infty}\left\{
		\frac{e^{-in\arccos(-b/a)}}{n}\right.\\ \left.\phantom{\frac{e^{n}}{n}}\times
		\left[\langle\bar{\vec{r}}|T_n(\tilde{H})|\vec{r}'\rangle+
		\langle\bar{\vec{r}}'|T_n(\tilde{H})|\vec{r}\rangle\right]\right\}^*.
	\end{multline}
The term of order $n=0$ disappears because $\langle\bar{\vec{r}}|T_0(\tilde{H})|\vec{r}'\rangle=\langle\bar{\vec{r}}|\vec{r}'\rangle=0$. Equation~(\ref{eq:gapequation}) must be solved self-consistently, with $H$ given by Eq.~(\ref{eq:block}).

\section{Double-tip tunneling}
\label{app:double-tip}

\begin{figure}[b]
\includegraphics[width=0.8\columnwidth]{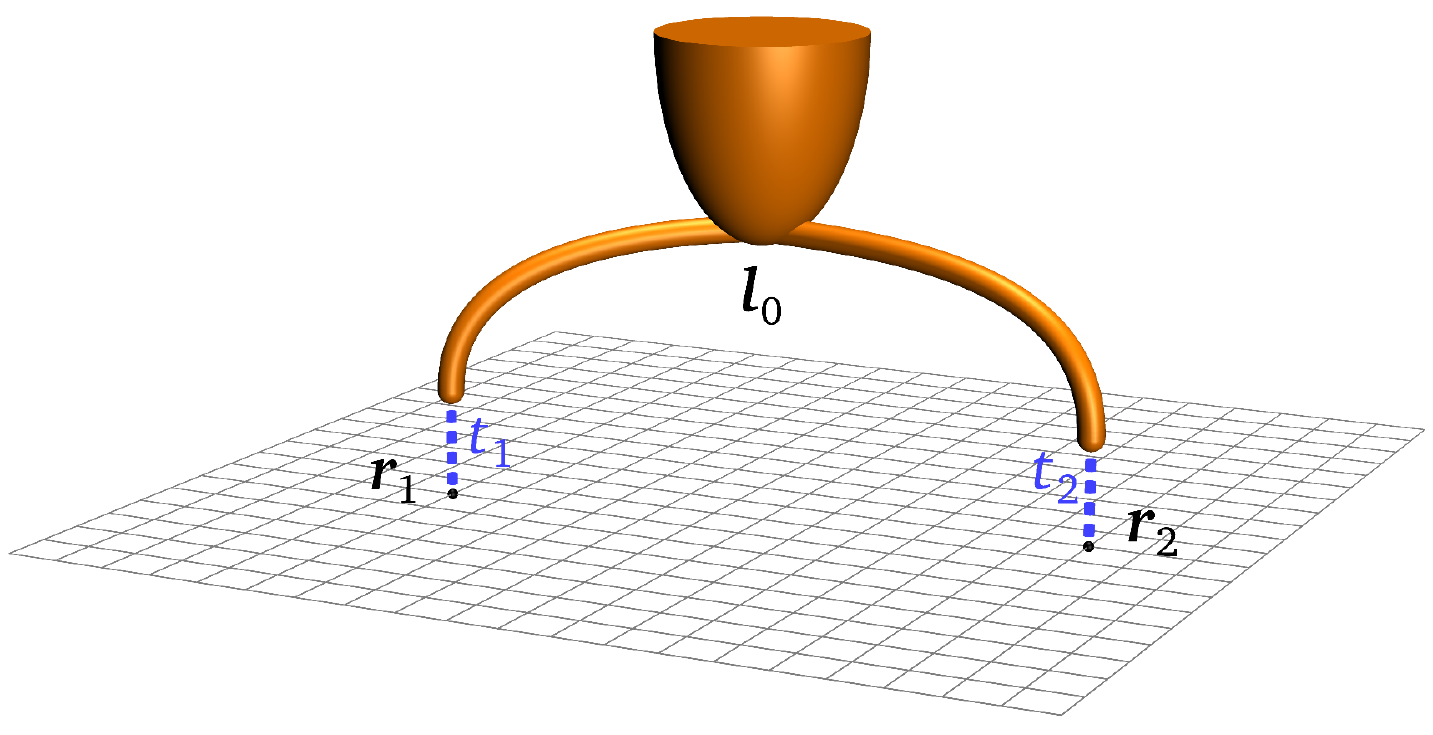}
\caption{\label{fig:fig-double}
Principle of double-tip tunneling to a common reservoir. The whole double-tip system (orange) is assumed to have the same density of states.
}
\end{figure}

Consider the usual electron tunneling problem, with two electrodes spanned by the vectors $\vec{l}$ (left) and $\vec{r}$ (right) and characterized by the single-particle retarded Green's functions $G(\vec{l},\vec{l}',\varepsilon)$ and $G(\vec{r},\vec{r}',\varepsilon)$, respectively. The two electrodes are coupled by a tunneling matrix element $t(\vec{l},\vec{r})$ and a chemical potential difference $eV$ is applied. The single-particle current at leading order in $t(\vec{l},\vec{r})$ is \cite{Todorov-1993}
	\begin{multline}\label{eq:Todorov}
		I(V)=\frac{2\pi e}{\hbar}\int_{-\infty}^{\infty}d\varepsilon\,[f(\varepsilon-eV)-f(\varepsilon)]\\
		\times\sum_{\vec{l}\vec{l}'\vec{r}\vec{r}'}\rho(\vec{l},\vec{l}',\varepsilon-eV)t^*(\vec{l}',\vec{r})
		\rho(\vec{r},\vec{r}',\varepsilon)t(\vec{l},\vec{r}').
	\end{multline}
$\rho(\vec{r},\vec{r}',\varepsilon)$ is the spectral function, which is related to the retarded Green's functions in the same way as in Eq.~(\ref{eq:rho11}). Although this formula was initially derived for quadratic Hamiltonians in the electrodes, it can be shown that the insertion of the interacting spectral functions in Eq.~(\ref{eq:Todorov}) correctly accounts for all correlations present in the electrodes, only neglecting correlations that are induced by the tunneling term \cite{Berthod-2011}. In the proposed setup (Fig.~\ref{fig:fig-double}), the tunneling matrix element vanishes unless $\vec{l}=\vec{l}_0$ and $\vec{r}=\vec{r}_1$ or $\vec{r}=\vec{r}_2$. We can write $t(\vec{l},\vec{r})=t_1\delta_{\vec{l}\vec{l}_0}\delta_{\vec{r}\vec{r}_1}+t_2\delta_{\vec{l}\vec{l}_0}\delta_{\vec{r}\vec{r}_2}$. We insert this into Eq.~(\ref{eq:Todorov}) and note that $\rho(\vec{l}_0,\vec{l}_0,\varepsilon-eV)$ is the LDOS in the left electrode representing the tip. We adopt the standard approximation and take it as a constant. Likewise, $\rho(\vec{r},\vec{r},\varepsilon)=(-1/\pi)\mathrm{Im}\,G(\vec{r},\vec{r},\varepsilon)=N(\vec{r},\varepsilon)$ is the LDOS in the right electrode. Since the bias $V$ only appears in one of the Fermi functions, we can differentiate and obtain the tunneling conductance as
	\begin{multline}\label{eq:two-1}
		\frac{dI}{dV}\propto\int_{-\infty}^{\infty}d\varepsilon\,[-f'(\varepsilon-eV)]
		\left\{t_1^2N(\vec{r}_1,\varepsilon)+t_2^2N(\vec{r}_2,\varepsilon)\right.\\\left.
		+t_1t_2\left({\textstyle-\frac{1}{\pi}}\right)\text{Im}\,[G(\vec{r}_1,\vec{r}_2,\varepsilon)
		+G(\vec{r}_2,\vec{r}_1,\varepsilon)]\right\}.
	\end{multline}
The first two terms in the curly braces give the parallel conductances associated with the two tips, proportional to the thermally-broadened LDOS at the points $\vec{r}_1$ and $\vec{r}_2$. The third term accounts for transport processes involving both tips. The latter is sensitive to the quasiparticle phase change between the points $\vec{r}_1$ and $\vec{r}_2$ and thus provides a correction to the conductance which can be positive or negative. If the amplitudes $t_1$ and $t_2$ are equal, it is seen that the relative change of conductance induced by the two-tip processes is given at zero temperature and for the energy $\varepsilon=eV$ by Eq.~(\ref{eq:conductance-ratio}).


\begin{thebibliography}{75}%
\makeatletter
\providecommand \@ifxundefined [1]{%
 \@ifx{#1\undefined}
}%
\providecommand \@ifnum [1]{%
 \ifnum #1\expandafter \@firstoftwo
 \else \expandafter \@secondoftwo
 \fi
}%
\providecommand \@ifx [1]{%
 \ifx #1\expandafter \@firstoftwo
 \else \expandafter \@secondoftwo
 \fi
}%
\providecommand \natexlab [1]{#1}%
\providecommand \enquote  [1]{``#1''}%
\providecommand \bibnamefont  [1]{#1}%
\providecommand \bibfnamefont [1]{#1}%
\providecommand \citenamefont [1]{#1}%
\providecommand \href@noop [0]{\@secondoftwo}%
\providecommand \href [0]{\begingroup \@sanitize@url \@href}%
\providecommand \@href[1]{\@@startlink{#1}\@@href}%
\providecommand \@@href[1]{\endgroup#1\@@endlink}%
\providecommand \@sanitize@url [0]{\catcode `\\12\catcode `\$12\catcode
  `\&12\catcode `\#12\catcode `\^12\catcode `\_12\catcode `\%12\relax}%
\providecommand \@@startlink[1]{}%
\providecommand \@@endlink[0]{}%
\providecommand \url  [0]{\begingroup\@sanitize@url \@url }%
\providecommand \@url [1]{\endgroup\@href {#1}{\urlprefix }}%
\providecommand \urlprefix  [0]{URL }%
\providecommand \Eprint [0]{\href }%
\providecommand \doibase [0]{http://dx.doi.org/}%
\providecommand \selectlanguage [0]{\@gobble}%
\providecommand \bibinfo  [0]{\@secondoftwo}%
\providecommand \bibfield  [0]{\@secondoftwo}%
\providecommand \translation [1]{[#1]}%
\providecommand \BibitemOpen [0]{}%
\providecommand \bibitemStop [0]{}%
\providecommand \bibitemNoStop [0]{.\EOS\space}%
\providecommand \EOS [0]{\spacefactor3000\relax}%
\providecommand \BibitemShut  [1]{\csname bibitem#1\endcsname}%
\let\auto@bib@innerbib\@empty
\bibitem [{\citenamefont {Hasan}\ and\ \citenamefont
  {Kane}(2010)}]{Hasan-2010}%
  \BibitemOpen
  \bibfield  {author} {\bibinfo {author} {\bibfnamefont {M.~Z.}\ \bibnamefont
  {Hasan}}\ and\ \bibinfo {author} {\bibfnamefont {C.~L.}\ \bibnamefont
  {Kane}},\ }\href {\doibase 10.1103/RevModPhys.82.3045} {\bibfield  {journal}
  {\bibinfo  {journal} {Rev. Mod. Phys.}\ }\textbf {\bibinfo {volume} {82}},\
  \bibinfo {pages} {3045} (\bibinfo {year} {2010})}\BibitemShut {NoStop}%
\bibitem [{\citenamefont {Volovik}(2003)}]{Volovik-2003}%
  \BibitemOpen
  \bibfield  {author} {\bibinfo {author} {\bibfnamefont {G.~E.}\ \bibnamefont
  {Volovik}},\ }\href {\doibase 10.1093/acprof:oso/9780199564842.003.0007}
  {\emph {\bibinfo {title} {The Universe in a Helium Droplet}}},\ International
  Series of Monographs on Physics\ (\bibinfo  {publisher} {Oxford University
  Press},\ \bibinfo {year} {2003})\BibitemShut {NoStop}%
\bibitem [{\citenamefont {Roy}(2010)}]{Roy-2010}%
  \BibitemOpen
  \bibfield  {author} {\bibinfo {author} {\bibfnamefont {R.}~\bibnamefont
  {Roy}},\ }\href {\doibase 10.1103/PhysRevLett.105.186401} {\bibfield
  {journal} {\bibinfo  {journal} {Phys. Rev. Lett.}\ }\textbf {\bibinfo
  {volume} {105}},\ \bibinfo {pages} {186401} (\bibinfo {year}
  {2010})}\BibitemShut {NoStop}%
\bibitem [{\citenamefont {Herbut}\ and\ \citenamefont
  {Lu}(2010)}]{Herbut-2010}%
  \BibitemOpen
  \bibfield  {author} {\bibinfo {author} {\bibfnamefont {I.~F.}\ \bibnamefont
  {Herbut}}\ and\ \bibinfo {author} {\bibfnamefont {C.-K.}\ \bibnamefont
  {Lu}},\ }\href {\doibase 10.1103/PhysRevB.82.125402} {\bibfield  {journal}
  {\bibinfo  {journal} {Phys. Rev. B}\ }\textbf {\bibinfo {volume} {82}},\
  \bibinfo {pages} {125402} (\bibinfo {year} {2010})}\BibitemShut {NoStop}%
\bibitem [{\citenamefont {Caroli}\ \emph {et~al.}(1964)\citenamefont {Caroli},
  \citenamefont {de~Gennes},\ and\ \citenamefont {Matricon}}]{Caroli-1964}%
  \BibitemOpen
  \bibfield  {author} {\bibinfo {author} {\bibfnamefont {C.}~\bibnamefont
  {Caroli}}, \bibinfo {author} {\bibfnamefont {P.~G.}\ \bibnamefont
  {de~Gennes}}, \ and\ \bibinfo {author} {\bibfnamefont {J.}~\bibnamefont
  {Matricon}},\ }\href {\doibase 10.1016/0031-9163(64)90375-0} {\bibfield
  {journal} {\bibinfo  {journal} {Phys. Lett.}\ }\textbf {\bibinfo {volume}
  {9}},\ \bibinfo {pages} {307} (\bibinfo {year} {1964})}\BibitemShut {NoStop}%
\bibitem [{\citenamefont {Gygi}\ and\ \citenamefont
  {Schl{\"u}ter}(1991)}]{Gygi-1991}%
  \BibitemOpen
  \bibfield  {author} {\bibinfo {author} {\bibfnamefont {F.}~\bibnamefont
  {Gygi}}\ and\ \bibinfo {author} {\bibfnamefont {M.}~\bibnamefont
  {Schl{\"u}ter}},\ }\href {\doibase 10.1103/PhysRevB.43.7609} {\bibfield
  {journal} {\bibinfo  {journal} {Phys. Rev. B}\ }\textbf {\bibinfo {volume}
  {43}},\ \bibinfo {pages} {7609} (\bibinfo {year} {1991})}\BibitemShut
  {NoStop}%
\bibitem [{\citenamefont {Hess}\ \emph {et~al.}(1989)\citenamefont {Hess},
  \citenamefont {Robinson}, \citenamefont {Dynes}, \citenamefont {Valles},\
  and\ \citenamefont {Waszczak}}]{Hess-1989}%
  \BibitemOpen
  \bibfield  {author} {\bibinfo {author} {\bibfnamefont {H.~F.}\ \bibnamefont
  {Hess}}, \bibinfo {author} {\bibfnamefont {R.~B.}\ \bibnamefont {Robinson}},
  \bibinfo {author} {\bibfnamefont {R.~C.}\ \bibnamefont {Dynes}}, \bibinfo
  {author} {\bibfnamefont {J.~M.}\ \bibnamefont {Valles}}, \ and\ \bibinfo
  {author} {\bibfnamefont {J.~V.}\ \bibnamefont {Waszczak}},\ }\href 
  {\doibase 10.1103/PhysRevLett.62.214} {\bibfield  {journal} {\bibinfo  {journal} {Phys.
  Rev. Lett.}\ }\textbf {\bibinfo {volume} {62}},\ \bibinfo {pages} {214}
  (\bibinfo {year} {1989})}\BibitemShut {NoStop}%
\bibitem [{\citenamefont {Guillam{\'o}n}\ \emph {et~al.}(2008)\citenamefont
  {Guillam{\'o}n}, \citenamefont {Suderow}, \citenamefont {Vieira},
  \citenamefont {Cario}, \citenamefont {Diener},\ and\ \citenamefont
  {Rodi{\`e}re}}]{Guillamon-2008a}%
  \BibitemOpen
  \bibfield  {author} {\bibinfo {author} {\bibfnamefont {I.}~\bibnamefont
  {Guillam{\'o}n}}, \bibinfo {author} {\bibfnamefont {H.}~\bibnamefont
  {Suderow}}, \bibinfo {author} {\bibfnamefont {S.}~\bibnamefont {Vieira}},
  \bibinfo {author} {\bibfnamefont {L.}~\bibnamefont {Cario}}, \bibinfo
  {author} {\bibfnamefont {P.}~\bibnamefont {Diener}}, \ and\ \bibinfo {author}
  {\bibfnamefont {P.}~\bibnamefont {Rodi{\`e}re}},\ }\href 
  {\doibase 10.1103/PhysRevLett.101.166407} {\bibfield  {journal} {\bibinfo  {journal}
  {Phys. Rev. Lett.}\ }\textbf {\bibinfo {volume} {101}},\ \bibinfo {pages}
  {166407} (\bibinfo {year} {2008})}\BibitemShut {NoStop}%
\bibitem [{\citenamefont {Maggio-Aprile}\ \emph {et~al.}(1995)\citenamefont
  {Maggio-Aprile}, \citenamefont {Renner}, \citenamefont {Erb}, \citenamefont
  {Walker},\ and\ \citenamefont {Fischer}}]{Maggio-Aprile-1995}%
  \BibitemOpen
  \bibfield  {author} {\bibinfo {author} {\bibfnamefont {I.}~\bibnamefont
  {Maggio-Aprile}}, \bibinfo {author} {\bibfnamefont {C.}~\bibnamefont
  {Renner}}, \bibinfo {author} {\bibfnamefont {A.}~\bibnamefont {Erb}},
  \bibinfo {author} {\bibfnamefont {E.}~\bibnamefont {Walker}}, \ and\ \bibinfo
  {author} {\bibfnamefont {{\O}.}~\bibnamefont {Fischer}},\ }\href 
  {\doibase 10.1103/PhysRevLett.75.2754} {\bibfield  {journal} {\bibinfo  {journal}
  {Phys. Rev. Lett.}\ }\textbf {\bibinfo {volume} {75}},\ \bibinfo {pages}
  {2754} (\bibinfo {year} {1995})}\BibitemShut {NoStop}%
\bibitem [{\citenamefont {Renner}\ \emph {et~al.}(1998)\citenamefont {Renner},
  \citenamefont {Revaz}, \citenamefont {Kadowaki}, \citenamefont
  {Maggio-Aprile},\ and\ \citenamefont {Fischer}}]{Renner-1998b}%
  \BibitemOpen
  \bibfield  {author} {\bibinfo {author} {\bibfnamefont {C.}~\bibnamefont
  {Renner}}, \bibinfo {author} {\bibfnamefont {B.}~\bibnamefont {Revaz}},
  \bibinfo {author} {\bibfnamefont {K.}~\bibnamefont {Kadowaki}}, \bibinfo
  {author} {\bibfnamefont {I.}~\bibnamefont {Maggio-Aprile}}, \ and\ \bibinfo
  {author} {\bibfnamefont {{\O}.}~\bibnamefont {Fischer}},\ }\href 
  {\doibase 10.1103/PhysRevLett.80.3606} {\bibfield  {journal} {\bibinfo  {journal}
  {Phys. Rev. Lett.}\ }\textbf {\bibinfo {volume} {80}},\ \bibinfo {pages}
  {3606} (\bibinfo {year} {1998})}\BibitemShut {NoStop}%
\bibitem [{\citenamefont {Hoogenboom}\ \emph {et~al.}(2000)\citenamefont
  {Hoogenboom}, \citenamefont {Renner}, \citenamefont {Revaz}, \citenamefont
  {Maggio-Aprile},\ and\ \citenamefont {Fischer}}]{Hoogenboom-2000a}%
  \BibitemOpen
  \bibfield  {author} {\bibinfo {author} {\bibfnamefont {B.~W.}\ \bibnamefont
  {Hoogenboom}}, \bibinfo {author} {\bibfnamefont {C.}~\bibnamefont {Renner}},
  \bibinfo {author} {\bibfnamefont {B.}~\bibnamefont {Revaz}}, \bibinfo
  {author} {\bibfnamefont {I.}~\bibnamefont {Maggio-Aprile}}, \ and\ \bibinfo
  {author} {\bibfnamefont {{\O}.}~\bibnamefont {Fischer}},\ }\href 
  {\doibase 10.1016/S0921-4534(99)00720-0} {\bibfield  {journal} {\bibinfo  {journal}
  {Physica C}\ }\textbf {\bibinfo {volume} {332}},\ \bibinfo {pages} {440}
  (\bibinfo {year} {2000})}\BibitemShut {NoStop}%
\bibitem [{\citenamefont {Pan}\ \emph {et~al.}(2000)\citenamefont {Pan},
  \citenamefont {Hudson}, \citenamefont {Gupta}, \citenamefont {Ng},
  \citenamefont {Eisaki}, \citenamefont {Uchida},\ and\ \citenamefont
  {Davis}}]{Pan-2000b}%
  \BibitemOpen
  \bibfield  {author} {\bibinfo {author} {\bibfnamefont {S.~H.}\ \bibnamefont
  {Pan}}, \bibinfo {author} {\bibfnamefont {E.~W.}\ \bibnamefont {Hudson}},
  \bibinfo {author} {\bibfnamefont {A.~K.}\ \bibnamefont {Gupta}}, \bibinfo
  {author} {\bibfnamefont {K.-W.}\ \bibnamefont {Ng}}, \bibinfo {author}
  {\bibfnamefont {H.}~\bibnamefont {Eisaki}}, \bibinfo {author} {\bibfnamefont
  {S.}~\bibnamefont {Uchida}}, \ and\ \bibinfo {author} {\bibfnamefont {J.~C.}\
  \bibnamefont {Davis}},\ }\href {\doibase 10.1103/PhysRevLett.85.1536}
  {\bibfield  {journal} {\bibinfo  {journal} {Phys. Rev. Lett.}\ }\textbf
  {\bibinfo {volume} {85}},\ \bibinfo {pages} {1536} (\bibinfo {year}
  {2000})}\BibitemShut {NoStop}%
\bibitem [{\citenamefont {Matsuba}\ \emph {et~al.}(2003)\citenamefont
  {Matsuba}, \citenamefont {Sakata}, \citenamefont {Kosugi}, \citenamefont
  {Nishimori},\ and\ \citenamefont {Nishida}}]{Matsuba-2003a}%
  \BibitemOpen
  \bibfield  {author} {\bibinfo {author} {\bibfnamefont {K.}~\bibnamefont
  {Matsuba}}, \bibinfo {author} {\bibfnamefont {H.}~\bibnamefont {Sakata}},
  \bibinfo {author} {\bibfnamefont {N.}~\bibnamefont {Kosugi}}, \bibinfo
  {author} {\bibfnamefont {H.}~\bibnamefont {Nishimori}}, \ and\ \bibinfo
  {author} {\bibfnamefont {N.}~\bibnamefont {Nishida}},\ }\href 
  {\doibase 10.1143/JPSJ.72.2153} {\bibfield  {journal} {\bibinfo  {journal} {J. Phys.
  Soc. Jpn.}\ }\textbf {\bibinfo {volume} {72}},\ \bibinfo {pages} {2153}
  (\bibinfo {year} {2003})}\BibitemShut {NoStop}%
\bibitem [{\citenamefont {Matsuba}\ \emph {et~al.}(2007)\citenamefont
  {Matsuba}, \citenamefont {Yoshizawa}, \citenamefont {Mochizuki},
  \citenamefont {Mochiku}, \citenamefont {Hirata},\ and\ \citenamefont
  {Nishida}}]{Matsuba-2007}%
  \BibitemOpen
  \bibfield  {author} {\bibinfo {author} {\bibfnamefont {K.}~\bibnamefont
  {Matsuba}}, \bibinfo {author} {\bibfnamefont {S.}~\bibnamefont {Yoshizawa}},
  \bibinfo {author} {\bibfnamefont {Y.}~\bibnamefont {Mochizuki}}, \bibinfo
  {author} {\bibfnamefont {T.}~\bibnamefont {Mochiku}}, \bibinfo {author}
  {\bibfnamefont {K.}~\bibnamefont {Hirata}}, \ and\ \bibinfo {author}
  {\bibfnamefont {N.}~\bibnamefont {Nishida}},\ }\href 
  {\doibase 10.1143/JPSJ.76.063704} {\bibfield  {journal} {\bibinfo  {journal} {J. Phys.
  Soc. Jpn.}\ }\textbf {\bibinfo {volume} {76}},\ \bibinfo {pages} {063704}
  (\bibinfo {year} {2007})}\BibitemShut {NoStop}%
\bibitem [{\citenamefont {Shibata}\ \emph {et~al.}(2003)\citenamefont
  {Shibata}, \citenamefont {Maki}, \citenamefont {Nishizaki},\ and\
  \citenamefont {Kobayashi}}]{Shibata-2003b}%
  \BibitemOpen
  \bibfield  {author} {\bibinfo {author} {\bibfnamefont {K.}~\bibnamefont
  {Shibata}}, \bibinfo {author} {\bibfnamefont {M.}~\bibnamefont {Maki}},
  \bibinfo {author} {\bibfnamefont {T.}~\bibnamefont {Nishizaki}}, \ and\
  \bibinfo {author} {\bibfnamefont {N.}~\bibnamefont {Kobayashi}},\ }\href
  {\doibase 10.1016/S0921-4534(03)01056-6} {\bibfield  {journal} {\bibinfo
  {journal} {Physica C}\ }\textbf {\bibinfo {volume} {392-396}},\ \bibinfo
  {pages} {323} (\bibinfo {year} {2003})}\BibitemShut {NoStop}%
\bibitem [{\citenamefont {Shibata}\ \emph {et~al.}(2010)\citenamefont
  {Shibata}, \citenamefont {Nishizaki}, \citenamefont {Maki},\ and\
  \citenamefont {Kobayashi}}]{Shibata-2010}%
  \BibitemOpen
  \bibfield  {author} {\bibinfo {author} {\bibfnamefont {K.}~\bibnamefont
  {Shibata}}, \bibinfo {author} {\bibfnamefont {T.}~\bibnamefont {Nishizaki}},
  \bibinfo {author} {\bibfnamefont {M.}~\bibnamefont {Maki}}, \ and\ \bibinfo
  {author} {\bibfnamefont {N.}~\bibnamefont {Kobayashi}},\ }\href 
  {\doibase 10.1088/0953-2048/23/8/085004} {\bibfield  {journal} {\bibinfo  {journal}
  {Supercond. Sci. Technol.}\ }\textbf {\bibinfo {volume} {23}},\ \bibinfo
  {pages} {085004} (\bibinfo {year} {2010})}\BibitemShut {NoStop}%
\bibitem [{\citenamefont {Levy}\ \emph {et~al.}(2005)\citenamefont {Levy},
  \citenamefont {Kugler}, \citenamefont {Manuel}, \citenamefont {Fischer},\
  and\ \citenamefont {Li}}]{Levy-2005}%
  \BibitemOpen
  \bibfield  {author} {\bibinfo {author} {\bibfnamefont {G.}~\bibnamefont
  {Levy}}, \bibinfo {author} {\bibfnamefont {M.}~\bibnamefont {Kugler}},
  \bibinfo {author} {\bibfnamefont {A.~A.}\ \bibnamefont {Manuel}}, \bibinfo
  {author} {\bibfnamefont {{\O}.}~\bibnamefont {Fischer}}, \ and\ \bibinfo
  {author} {\bibfnamefont {M.}~\bibnamefont {Li}},\ }\href 
  {\doibase 10.1103/PhysRevLett.95.257005} {\bibfield  {journal} {\bibinfo  {journal}
  {Phys. Rev. Lett.}\ }\textbf {\bibinfo {volume} {95}},\ \bibinfo {pages}
  {257005} (\bibinfo {year} {2005})}\BibitemShut {NoStop}%
\bibitem [{\citenamefont {Yoshizawa}\ \emph {et~al.}(2013)\citenamefont
  {Yoshizawa}, \citenamefont {Koseki}, \citenamefont {Matsuba}, \citenamefont
  {Mochiku}, \citenamefont {Hirata},\ and\ \citenamefont
  {Nishida}}]{Yoshizawa-2013}%
  \BibitemOpen
  \bibfield  {author} {\bibinfo {author} {\bibfnamefont {S.}~\bibnamefont
  {Yoshizawa}}, \bibinfo {author} {\bibfnamefont {T.}~\bibnamefont {Koseki}},
  \bibinfo {author} {\bibfnamefont {K.}~\bibnamefont {Matsuba}}, \bibinfo
  {author} {\bibfnamefont {T.}~\bibnamefont {Mochiku}}, \bibinfo {author}
  {\bibfnamefont {K.}~\bibnamefont {Hirata}}, \ and\ \bibinfo {author}
  {\bibfnamefont {N.}~\bibnamefont {Nishida}},\ }\href 
  {\doibase 10.7566/JPSJ.82.083706} {\bibfield  {journal} {\bibinfo  {journal} {J. Phys.
  Soc. Jpn.}\ }\textbf {\bibinfo {volume} {82}},\ \bibinfo {pages} {083706}
  (\bibinfo {year} {2013})}\BibitemShut {NoStop}%
\bibitem [{\citenamefont {Wang}\ and\ \citenamefont
  {MacDonald}(1995)}]{Wang-1995}%
  \BibitemOpen
  \bibfield  {author} {\bibinfo {author} {\bibfnamefont {Y.}~\bibnamefont
  {Wang}}\ and\ \bibinfo {author} {\bibfnamefont {A.~H.}\ \bibnamefont
  {MacDonald}},\ }\href {\doibase 10.1103/PhysRevB.52.R3876} {\bibfield
  {journal} {\bibinfo  {journal} {Phys. Rev. B}\ }\textbf {\bibinfo {volume}
  {52}},\ \bibinfo {pages} {R3876} (\bibinfo {year} {1995})}\BibitemShut
  {NoStop}%
\bibitem [{\citenamefont {Franz}\ and\ \citenamefont
  {Te{\v{s}}anovi{\'{c}}}(1998)}]{Franz-1998b}%
  \BibitemOpen
  \bibfield  {author} {\bibinfo {author} {\bibfnamefont {M.}~\bibnamefont
  {Franz}}\ and\ \bibinfo {author} {\bibfnamefont {Z.}~\bibnamefont
  {Te{\v{s}}anovi{\'{c}}}},\ }\href {\doibase 10.1103/PhysRevLett.80.4763}
  {\bibfield  {journal} {\bibinfo  {journal} {Phys. Rev. Lett.}\ }\textbf
  {\bibinfo {volume} {80}},\ \bibinfo {pages} {4763} (\bibinfo {year}
  {1998})}\BibitemShut {NoStop}%
\bibitem [{\citenamefont {Yasui}\ and\ \citenamefont
  {Kita}(1999)}]{Yasui-1999}%
  \BibitemOpen
  \bibfield  {author} {\bibinfo {author} {\bibfnamefont {K.}~\bibnamefont
  {Yasui}}\ and\ \bibinfo {author} {\bibfnamefont {T.}~\bibnamefont {Kita}},\
  }\href {\doibase 10.1103/PhysRevLett.83.4168} {\bibfield  {journal} {\bibinfo
   {journal} {Phys. Rev. Lett.}\ }\textbf {\bibinfo {volume} {83}},\ \bibinfo
  {pages} {4168} (\bibinfo {year} {1999})}\BibitemShut {NoStop}%
\bibitem [{\citenamefont {Arovas}\ \emph {et~al.}(1997)\citenamefont {Arovas},
  \citenamefont {Berlinsky}, \citenamefont {Kallin},\ and\ \citenamefont
  {Zhang}}]{Arovas-1997}%
  \BibitemOpen
  \bibfield  {author} {\bibinfo {author} {\bibfnamefont {D.~P.}\ \bibnamefont
  {Arovas}}, \bibinfo {author} {\bibfnamefont {A.~J.}\ \bibnamefont
  {Berlinsky}}, \bibinfo {author} {\bibfnamefont {C.}~\bibnamefont {Kallin}}, \
  and\ \bibinfo {author} {\bibfnamefont {S.-C.}\ \bibnamefont {Zhang}},\ }\href
  {\doibase 10.1103/PhysRevLett.79.2871} {\bibfield  {journal} {\bibinfo
  {journal} {Phys. Rev. Lett.}\ }\textbf {\bibinfo {volume} {79}},\ \bibinfo
  {pages} {2871} (\bibinfo {year} {1997})}\BibitemShut {NoStop}%
\bibitem [{\citenamefont {Himeda}\ \emph {et~al.}(1997)\citenamefont {Himeda},
  \citenamefont {Ogata}, \citenamefont {Tanaka},\ and\ \citenamefont
  {Kashiwaya}}]{Himeda-1997}%
  \BibitemOpen
  \bibfield  {author} {\bibinfo {author} {\bibfnamefont {A.}~\bibnamefont
  {Himeda}}, \bibinfo {author} {\bibfnamefont {M.}~\bibnamefont {Ogata}},
  \bibinfo {author} {\bibfnamefont {Y.}~\bibnamefont {Tanaka}}, \ and\ \bibinfo
  {author} {\bibfnamefont {S.}~\bibnamefont {Kashiwaya}},\ }\href 
  {\doibase 10.1143/JPSJ.66.3367} {\bibfield  {journal} {\bibinfo  {journal} {J. Phys.
  Soc. Jpn.}\ }\textbf {\bibinfo {volume} {66}},\ \bibinfo {pages} {3367}
  (\bibinfo {year} {1997})}\BibitemShut {NoStop}%
\bibitem [{\citenamefont {Andersen}\ \emph {et~al.}(2000)\citenamefont
  {Andersen}, \citenamefont {Bruus},\ and\ \citenamefont
  {Hedeg{\aa}rd}}]{Andersen-2000}%
  \BibitemOpen
  \bibfield  {author} {\bibinfo {author} {\bibfnamefont {B.~M.}\ \bibnamefont
  {Andersen}}, \bibinfo {author} {\bibfnamefont {H.}~\bibnamefont {Bruus}}, \
  and\ \bibinfo {author} {\bibfnamefont {P.}~\bibnamefont {Hedeg{\aa}rd}},\
  }\href {\doibase 10.1103/PhysRevB.61.6298} {\bibfield  {journal} {\bibinfo
  {journal} {Phys. Rev. B}\ }\textbf {\bibinfo {volume} {61}},\ \bibinfo
  {pages} {6298} (\bibinfo {year} {2000})}\BibitemShut {NoStop}%
\bibitem [{\citenamefont {Wu}\ \emph {et~al.}(2000)\citenamefont {Wu},
  \citenamefont {Xiang},\ and\ \citenamefont {Su}}]{Wu-2000}%
  \BibitemOpen
  \bibfield  {author} {\bibinfo {author} {\bibfnamefont {C.}~\bibnamefont
  {Wu}}, \bibinfo {author} {\bibfnamefont {T.}~\bibnamefont {Xiang}}, \ and\
  \bibinfo {author} {\bibfnamefont {Z.-B.}\ \bibnamefont {Su}},\ }\href
  {\doibase 10.1103/PhysRevB.62.14427} {\bibfield  {journal} {\bibinfo
  {journal} {Phys. Rev. B}\ }\textbf {\bibinfo {volume} {62}},\ \bibinfo
  {pages} {14427} (\bibinfo {year} {2000})}\BibitemShut {NoStop}%
\bibitem [{\citenamefont {Kishine}\ \emph {et~al.}(2001)\citenamefont
  {Kishine}, \citenamefont {Lee},\ and\ \citenamefont {Wen}}]{Kishine-2001}%
  \BibitemOpen
  \bibfield  {author} {\bibinfo {author} {\bibfnamefont {J.-i.}\ \bibnamefont
  {Kishine}}, \bibinfo {author} {\bibfnamefont {P.~A.}\ \bibnamefont {Lee}}, \
  and\ \bibinfo {author} {\bibfnamefont {X.-G.}\ \bibnamefont {Wen}},\ }\href
  {\doibase 10.1103/PhysRevLett.86.5365} {\bibfield  {journal} {\bibinfo
  {journal} {Phys. Rev. Lett.}\ }\textbf {\bibinfo {volume} {86}},\ \bibinfo
  {pages} {5365} (\bibinfo {year} {2001})}\BibitemShut {NoStop}%
\bibitem [{\citenamefont {Berthod}\ and\ \citenamefont
  {Giovannini}(2001)}]{Berthod-2001b}%
  \BibitemOpen
  \bibfield  {author} {\bibinfo {author} {\bibfnamefont {C.}~\bibnamefont
  {Berthod}}\ and\ \bibinfo {author} {\bibfnamefont {B.}~\bibnamefont
  {Giovannini}},\ }\href {\doibase 10.1103/PhysRevLett.87.277002} {\bibfield
  {journal} {\bibinfo  {journal} {Phys. Rev. Lett.}\ }\textbf {\bibinfo
  {volume} {87}},\ \bibinfo {pages} {277002} (\bibinfo {year}
  {2001})}\BibitemShut {NoStop}%
\bibitem [{\citenamefont {Zhu}\ and\ \citenamefont {Ting}(2001)}]{Zhu-2001a}%
  \BibitemOpen
  \bibfield  {author} {\bibinfo {author} {\bibfnamefont {J.-X.}\ \bibnamefont
  {Zhu}}\ and\ \bibinfo {author} {\bibfnamefont {C.~S.}\ \bibnamefont {Ting}},\
  }\href {\doibase 10.1103/PhysRevLett.87.147002} {\bibfield  {journal}
  {\bibinfo  {journal} {Phys. Rev. Lett.}\ }\textbf {\bibinfo {volume} {87}},\
  \bibinfo {pages} {147002} (\bibinfo {year} {2001})}\BibitemShut {NoStop}%
\bibitem [{\citenamefont {Chen}\ and\ \citenamefont {Ting}(2002)}]{Chen-2002b}%
  \BibitemOpen
  \bibfield  {author} {\bibinfo {author} {\bibfnamefont {Y.}~\bibnamefont
  {Chen}}\ and\ \bibinfo {author} {\bibfnamefont {C.~S.}\ \bibnamefont
  {Ting}},\ }\href {\doibase 10.1103/PhysRevB.65.180513} {\bibfield  {journal}
  {\bibinfo  {journal} {Phys. Rev. B}\ }\textbf {\bibinfo {volume} {65}},\
  \bibinfo {pages} {180513(R)} (\bibinfo {year} {2002})}\BibitemShut {NoStop}%
\bibitem [{\citenamefont {Ma{\'{s}}ka}\ and\ \citenamefont
  {Mierzejewski}(2003)}]{Maska-2003}%
  \BibitemOpen
  \bibfield  {author} {\bibinfo {author} {\bibfnamefont {M.~M.}\ \bibnamefont
  {Ma{\'{s}}ka}}\ and\ \bibinfo {author} {\bibfnamefont {M.}~\bibnamefont
  {Mierzejewski}},\ }\href {\doibase 10.1103/PhysRevB.68.024513} {\bibfield
  {journal} {\bibinfo  {journal} {Phys. Rev. B}\ }\textbf {\bibinfo {volume}
  {68}},\ \bibinfo {pages} {024513} (\bibinfo {year} {2003})}\BibitemShut
  {NoStop}%
\bibitem [{\citenamefont {Tsuchiura}\ \emph {et~al.}(2003)\citenamefont
  {Tsuchiura}, \citenamefont {Ogata}, \citenamefont {Tanaka},\ and\
  \citenamefont {Kashiwaya}}]{Tsuchiura-2003}%
  \BibitemOpen
  \bibfield  {author} {\bibinfo {author} {\bibfnamefont {H.}~\bibnamefont
  {Tsuchiura}}, \bibinfo {author} {\bibfnamefont {M.}~\bibnamefont {Ogata}},
  \bibinfo {author} {\bibfnamefont {Y.}~\bibnamefont {Tanaka}}, \ and\ \bibinfo
  {author} {\bibfnamefont {S.}~\bibnamefont {Kashiwaya}},\ }\href 
  {\doibase 10.1103/PhysRevB.68.012509} {\bibfield  {journal} {\bibinfo  {journal} {Phys.
  Rev. B}\ }\textbf {\bibinfo {volume} {68}},\ \bibinfo {pages} {012509}
  (\bibinfo {year} {2003})}\BibitemShut {NoStop}%
\bibitem [{\citenamefont {Takigawa}\ \emph {et~al.}(2003)\citenamefont
  {Takigawa}, \citenamefont {Ichioka},\ and\ \citenamefont
  {Machida}}]{Takigawa-2003}%
  \BibitemOpen
  \bibfield  {author} {\bibinfo {author} {\bibfnamefont {M.}~\bibnamefont
  {Takigawa}}, \bibinfo {author} {\bibfnamefont {M.}~\bibnamefont {Ichioka}}, \
  and\ \bibinfo {author} {\bibfnamefont {K.}~\bibnamefont {Machida}},\ }\href
  {\doibase 10.1103/PhysRevLett.90.047001} {\bibfield  {journal} {\bibinfo
  {journal} {Phys. Rev. Lett.}\ }\textbf {\bibinfo {volume} {90}},\ \bibinfo
  {pages} {047001} (\bibinfo {year} {2003})}\BibitemShut {NoStop}%
\bibitem [{\citenamefont {Takigawa}\ \emph {et~al.}(2004)\citenamefont
  {Takigawa}, \citenamefont {Ichioka},\ and\ \citenamefont
  {Machida}}]{Takigawa-2004}%
  \BibitemOpen
  \bibfield  {author} {\bibinfo {author} {\bibfnamefont {M.}~\bibnamefont
  {Takigawa}}, \bibinfo {author} {\bibfnamefont {M.}~\bibnamefont {Ichioka}}, \
  and\ \bibinfo {author} {\bibfnamefont {K.}~\bibnamefont {Machida}},\ }\href
  {\doibase 10.1143/JPSJ.73.450} {\bibfield  {journal} {\bibinfo  {journal} {J.
  Phys. Soc. Jpn.}\ }\textbf {\bibinfo {volume} {73}},\ \bibinfo {pages} {450}
  (\bibinfo {year} {2004})}\BibitemShut {NoStop}%
\bibitem [{\citenamefont {Fogelstr{\"o}m}(2011)}]{Fogelstrom-2011}%
  \BibitemOpen
  \bibfield  {author} {\bibinfo {author} {\bibfnamefont {M.}~\bibnamefont
  {Fogelstr{\"o}m}},\ }\href {\doibase 10.1103/PhysRevB.84.064530} {\bibfield
  {journal} {\bibinfo  {journal} {Phys. Rev. B}\ }\textbf {\bibinfo {volume}
  {84}},\ \bibinfo {pages} {064530} (\bibinfo {year} {2011})}\BibitemShut
  {NoStop}%
\bibitem [{\citenamefont {Song}\ \emph {et~al.}(2011)\citenamefont {Song},
  \citenamefont {Wang}, \citenamefont {Cheng}, \citenamefont {Jiang},
  \citenamefont {Li}, \citenamefont {Zhang}, \citenamefont {Li}, \citenamefont
  {He}, \citenamefont {Wang}, \citenamefont {Jia}, \citenamefont {Hung},
  \citenamefont {Wu}, \citenamefont {Ma}, \citenamefont {Chen},\ and\
  \citenamefont {Xue}}]{Song-2011}%
  \BibitemOpen
  \bibfield  {author} {\bibinfo {author} {\bibfnamefont {C.-L.}\ \bibnamefont
  {Song}}, \bibinfo {author} {\bibfnamefont {Y.-L.}\ \bibnamefont {Wang}},
  \bibinfo {author} {\bibfnamefont {P.}~\bibnamefont {Cheng}}, \bibinfo
  {author} {\bibfnamefont {Y.-P.}\ \bibnamefont {Jiang}}, \bibinfo {author}
  {\bibfnamefont {W.}~\bibnamefont {Li}}, \bibinfo {author} {\bibfnamefont
  {T.}~\bibnamefont {Zhang}}, \bibinfo {author} {\bibfnamefont
  {Z.}~\bibnamefont {Li}}, \bibinfo {author} {\bibfnamefont {K.}~\bibnamefont
  {He}}, \bibinfo {author} {\bibfnamefont {L.}~\bibnamefont {Wang}}, \bibinfo
  {author} {\bibfnamefont {J.-F.}\ \bibnamefont {Jia}}, \bibinfo {author}
  {\bibfnamefont {H.-H.}\ \bibnamefont {Hung}}, \bibinfo {author}
  {\bibfnamefont {C.}~\bibnamefont {Wu}}, \bibinfo {author} {\bibfnamefont
  {X.}~\bibnamefont {Ma}}, \bibinfo {author} {\bibfnamefont {X.}~\bibnamefont
  {Chen}}, \ and\ \bibinfo {author} {\bibfnamefont {Q.-K.}\ \bibnamefont
  {Xue}},\ }\href {\doibase 10.1126/science.1202226} {\bibfield  {journal}
  {\bibinfo  {journal} {Science}\ }\textbf {\bibinfo {volume} {332}},\ \bibinfo
  {pages} {1410} (\bibinfo {year} {2011})}\BibitemShut {NoStop}%
\bibitem [{\citenamefont {Shan}\ \emph {et~al.}(2011)\citenamefont {Shan},
  \citenamefont {Wang}, \citenamefont {Shen}, \citenamefont {Zeng},
  \citenamefont {Huang}, \citenamefont {Li}, \citenamefont {Wang},
  \citenamefont {Yang}, \citenamefont {Ren}, \citenamefont {Wang},
  \citenamefont {Pan},\ and\ \citenamefont {Wen}}]{Shan-2011}%
  \BibitemOpen
  \bibfield  {author} {\bibinfo {author} {\bibfnamefont {L.}~\bibnamefont
  {Shan}}, \bibinfo {author} {\bibfnamefont {Y.-L.}\ \bibnamefont {Wang}},
  \bibinfo {author} {\bibfnamefont {B.}~\bibnamefont {Shen}}, \bibinfo {author}
  {\bibfnamefont {B.}~\bibnamefont {Zeng}}, \bibinfo {author} {\bibfnamefont
  {Y.}~\bibnamefont {Huang}}, \bibinfo {author} {\bibfnamefont
  {A.}~\bibnamefont {Li}}, \bibinfo {author} {\bibfnamefont {D.}~\bibnamefont
  {Wang}}, \bibinfo {author} {\bibfnamefont {H.}~\bibnamefont {Yang}}, \bibinfo
  {author} {\bibfnamefont {C.}~\bibnamefont {Ren}}, \bibinfo {author}
  {\bibfnamefont {Q.-H.}\ \bibnamefont {Wang}}, \bibinfo {author}
  {\bibfnamefont {S.~H.}\ \bibnamefont {Pan}}, \ and\ \bibinfo {author}
  {\bibfnamefont {H.-H.}\ \bibnamefont {Wen}},\ }\href 
  {\doibase 10.1038/nphys1908} {\bibfield  {journal} {\bibinfo  {journal} {Nat. Phys.}\
  }\textbf {\bibinfo {volume} {7}},\ \bibinfo {pages} {325} (\bibinfo {year}
  {2011})}\BibitemShut {NoStop}%
\bibitem [{\citenamefont {Hanaguri}\ \emph {et~al.}(2012)\citenamefont
  {Hanaguri}, \citenamefont {Kitagawa}, \citenamefont {Matsubayashi},
  \citenamefont {Mazaki}, \citenamefont {Uwatoko},\ and\ \citenamefont
  {Takagi}}]{Hanaguri-2012}%
  \BibitemOpen
  \bibfield  {author} {\bibinfo {author} {\bibfnamefont {T.}~\bibnamefont
  {Hanaguri}}, \bibinfo {author} {\bibfnamefont {K.}~\bibnamefont {Kitagawa}},
  \bibinfo {author} {\bibfnamefont {K.}~\bibnamefont {Matsubayashi}}, \bibinfo
  {author} {\bibfnamefont {Y.}~\bibnamefont {Mazaki}}, \bibinfo {author}
  {\bibfnamefont {Y.}~\bibnamefont {Uwatoko}}, \ and\ \bibinfo {author}
  {\bibfnamefont {H.}~\bibnamefont {Takagi}},\ }\href 
  {\doibase 10.1103/PhysRevB.85.214505} {\bibfield  {journal} {\bibinfo  {journal} {Phys.
  Rev. B}\ }\textbf {\bibinfo {volume} {85}},\ \bibinfo {pages} {214505}
  (\bibinfo {year} {2012})}\BibitemShut {NoStop}%
\bibitem [{\citenamefont {Yin}\ \emph {et~al.}(2009)\citenamefont {Yin},
  \citenamefont {Zech}, \citenamefont {Williams}, \citenamefont {Wang},
  \citenamefont {Wu}, \citenamefont {Chen},\ and\ \citenamefont
  {Hoffman}}]{Yin-2009}%
  \BibitemOpen
  \bibfield  {author} {\bibinfo {author} {\bibfnamefont {Y.}~\bibnamefont
  {Yin}}, \bibinfo {author} {\bibfnamefont {M.}~\bibnamefont {Zech}}, \bibinfo
  {author} {\bibfnamefont {T.~L.}\ \bibnamefont {Williams}}, \bibinfo {author}
  {\bibfnamefont {X.~F.}\ \bibnamefont {Wang}}, \bibinfo {author}
  {\bibfnamefont {G.}~\bibnamefont {Wu}}, \bibinfo {author} {\bibfnamefont
  {X.~H.}\ \bibnamefont {Chen}}, \ and\ \bibinfo {author} {\bibfnamefont
  {J.~E.}\ \bibnamefont {Hoffman}},\ }\href 
  {\doibase 10.1103/PhysRevLett.102.097002} {\bibfield  {journal} {\bibinfo  {journal}
  {Phys. Rev. Lett.}\ }\textbf {\bibinfo {volume} {102}},\ \bibinfo {pages}
  {097002} (\bibinfo {year} {2009})}\BibitemShut {NoStop}%
\bibitem [{\citenamefont {Renner}\ \emph {et~al.}(1991)\citenamefont {Renner},
  \citenamefont {Kent}, \citenamefont {Niedermann}, \citenamefont {Fischer},\
  and\ \citenamefont {L{\'e}vy}}]{Renner-1991}%
  \BibitemOpen
  \bibfield  {author} {\bibinfo {author} {\bibfnamefont {C.}~\bibnamefont
  {Renner}}, \bibinfo {author} {\bibfnamefont {A.~D.}\ \bibnamefont {Kent}},
  \bibinfo {author} {\bibfnamefont {P.}~\bibnamefont {Niedermann}}, \bibinfo
  {author} {\bibfnamefont {{\O}.}~\bibnamefont {Fischer}}, \ and\ \bibinfo
  {author} {\bibfnamefont {F.}~\bibnamefont {L{\'e}vy}},\ }\href 
  {\doibase 10.1103/PhysRevLett.67.1650} {\bibfield  {journal} {\bibinfo  {journal}
  {Phys. Rev. Lett.}\ }\textbf {\bibinfo {volume} {67}},\ \bibinfo {pages}
  {1650} (\bibinfo {year} {1991})}\BibitemShut {NoStop}%
\bibitem [{\citenamefont {Eskildsen}\ \emph {et~al.}(2002)\citenamefont
  {Eskildsen}, \citenamefont {Kugler}, \citenamefont {Tanaka}, \citenamefont
  {Jun}, \citenamefont {Kazakov}, \citenamefont {Karpinski},\ and\
  \citenamefont {Fischer}}]{Eskildsen-2002}%
  \BibitemOpen
  \bibfield  {author} {\bibinfo {author} {\bibfnamefont {M.~R.}\ \bibnamefont
  {Eskildsen}}, \bibinfo {author} {\bibfnamefont {M.}~\bibnamefont {Kugler}},
  \bibinfo {author} {\bibfnamefont {S.}~\bibnamefont {Tanaka}}, \bibinfo
  {author} {\bibfnamefont {J.}~\bibnamefont {Jun}}, \bibinfo {author}
  {\bibfnamefont {S.~M.}\ \bibnamefont {Kazakov}}, \bibinfo {author}
  {\bibfnamefont {J.}~\bibnamefont {Karpinski}}, \ and\ \bibinfo {author}
  {\bibfnamefont {{\O}.}~\bibnamefont {Fischer}},\ }\href 
  {\doibase 10.1103/PhysRevLett.89.187003} {\bibfield  {journal} {\bibinfo  {journal}
  {Phys. Rev. Lett.}\ }\textbf {\bibinfo {volume} {89}},\ \bibinfo {pages}
  {187003} (\bibinfo {year} {2002})}\BibitemShut {NoStop}%
\bibitem [{\citenamefont {Fischer}\ \emph {et~al.}(2007)\citenamefont
  {Fischer}, \citenamefont {Kugler}, \citenamefont {Maggio-Aprile},
  \citenamefont {Berthod},\ and\ \citenamefont {Renner}}]{Fischer-2007}%
  \BibitemOpen
  \bibfield  {author} {\bibinfo {author} {\bibfnamefont {{\O}.}~\bibnamefont
  {Fischer}}, \bibinfo {author} {\bibfnamefont {M.}~\bibnamefont {Kugler}},
  \bibinfo {author} {\bibfnamefont {I.}~\bibnamefont {Maggio-Aprile}}, \bibinfo
  {author} {\bibfnamefont {C.}~\bibnamefont {Berthod}}, \ and\ \bibinfo
  {author} {\bibfnamefont {C.}~\bibnamefont {Renner}},\ }\href 
  {\doibase 10.1103/RevModPhys.79.353} {\bibfield  {journal} {\bibinfo  {journal} {Rev.
  Mod. Phys.}\ }\textbf {\bibinfo {volume} {79}},\ \bibinfo {pages} {353}
  (\bibinfo {year} {2007})}\BibitemShut {NoStop}%
\bibitem [{\citenamefont {Eschrig}\ and\ \citenamefont
  {Norman}(2000)}]{Eschrig-2000}%
  \BibitemOpen
  \bibfield  {author} {\bibinfo {author} {\bibfnamefont {M.}~\bibnamefont
  {Eschrig}}\ and\ \bibinfo {author} {\bibfnamefont {M.~R.}\ \bibnamefont
  {Norman}},\ }\href {\doibase 10.1103/PhysRevLett.85.3261} {\bibfield
  {journal} {\bibinfo  {journal} {Phys. Rev. Lett.}\ }\textbf {\bibinfo
  {volume} {85}},\ \bibinfo {pages} {3261} (\bibinfo {year}
  {2000})}\BibitemShut {NoStop}%
\bibitem [{\citenamefont {Berthod}\ \emph {et~al.}(2013)\citenamefont
  {Berthod}, \citenamefont {Fasano}, \citenamefont {Maggio-Aprile},
  \citenamefont {Piriou}, \citenamefont {Giannini}, \citenamefont {Levy~de
  Castro},\ and\ \citenamefont {Fischer}}]{Berthod-2013}%
  \BibitemOpen
  \bibfield  {author} {\bibinfo {author} {\bibfnamefont {C.}~\bibnamefont
  {Berthod}}, \bibinfo {author} {\bibfnamefont {Y.}~\bibnamefont {Fasano}},
  \bibinfo {author} {\bibfnamefont {I.}~\bibnamefont {Maggio-Aprile}}, \bibinfo
  {author} {\bibfnamefont {A.}~\bibnamefont {Piriou}}, \bibinfo {author}
  {\bibfnamefont {E.}~\bibnamefont {Giannini}}, \bibinfo {author}
  {\bibfnamefont {G.}~\bibnamefont {Levy~de Castro}}, \ and\ \bibinfo {author}
  {\bibfnamefont {{\O}.}~\bibnamefont {Fischer}},\ }\href 
  {\doibase 10.1103/PhysRevB.88.014528} {\bibfield  {journal} {\bibinfo  {journal} {Phys.
  Rev. B}\ }\textbf {\bibinfo {volume} {88}},\ \bibinfo {pages} {014528}
  (\bibinfo {year} {2013})}\BibitemShut {NoStop}%
\bibitem [{\citenamefont {Vaknin}\ \emph {et~al.}(2000)\citenamefont {Vaknin},
  \citenamefont {Zarestky},\ and\ \citenamefont {Miller}}]{Vaknin-2000}%
  \BibitemOpen
  \bibfield  {author} {\bibinfo {author} {\bibfnamefont {D.}~\bibnamefont
  {Vaknin}}, \bibinfo {author} {\bibfnamefont {J.~L.}\ \bibnamefont
  {Zarestky}}, \ and\ \bibinfo {author} {\bibfnamefont {L.~L.}\ \bibnamefont
  {Miller}},\ }\href {\doibase 10.1016/S0921-4534(99)00558-4} {\bibfield
  {journal} {\bibinfo  {journal} {Physica C}\ }\textbf {\bibinfo {volume}
  {329}},\ \bibinfo {pages} {109} (\bibinfo {year} {2000})}\BibitemShut
  {NoStop}%
\bibitem [{\citenamefont {Lake}\ \emph {et~al.}(2001)\citenamefont {Lake},
  \citenamefont {Aeppli}, \citenamefont {Clausen}, \citenamefont {McMorrow},
  \citenamefont {Lefmann}, \citenamefont {Hussey}, \citenamefont
  {Mangkorntong}, \citenamefont {Nohara}, \citenamefont {Takagi}, \citenamefont
  {Mason},\ and\ \citenamefont {Schr{\"o}der}}]{Lake-2001}%
  \BibitemOpen
  \bibfield  {author} {\bibinfo {author} {\bibfnamefont {B.}~\bibnamefont
  {Lake}}, \bibinfo {author} {\bibfnamefont {G.}~\bibnamefont {Aeppli}},
  \bibinfo {author} {\bibfnamefont {K.~N.}\ \bibnamefont {Clausen}}, \bibinfo
  {author} {\bibfnamefont {D.~F.}\ \bibnamefont {McMorrow}}, \bibinfo {author}
  {\bibfnamefont {K.}~\bibnamefont {Lefmann}}, \bibinfo {author} {\bibfnamefont
  {N.~E.}\ \bibnamefont {Hussey}}, \bibinfo {author} {\bibfnamefont
  {N.}~\bibnamefont {Mangkorntong}}, \bibinfo {author} {\bibfnamefont
  {M.}~\bibnamefont {Nohara}}, \bibinfo {author} {\bibfnamefont
  {H.}~\bibnamefont {Takagi}}, \bibinfo {author} {\bibfnamefont {T.~E.}\
  \bibnamefont {Mason}}, \ and\ \bibinfo {author} {\bibfnamefont
  {A.}~\bibnamefont {Schr{\"o}der}},\ }\href {\doibase 10.1126/science.1056986}
  {\bibfield  {journal} {\bibinfo  {journal} {Science}\ }\textbf {\bibinfo
  {volume} {291}},\ \bibinfo {pages} {1759} (\bibinfo {year}
  {2001})}\BibitemShut {NoStop}%
\bibitem [{\citenamefont {Lake}\ \emph {et~al.}(2002)\citenamefont {Lake},
  \citenamefont {R{\o}nnow}, \citenamefont {Christensen}, \citenamefont
  {Aeppli}, \citenamefont {Lefmann}, \citenamefont {McMorrow}, \citenamefont
  {Vorderwisch}, \citenamefont {Smeibidl}, \citenamefont {Mangkorntong},
  \citenamefont {Sasagawa}, \citenamefont {Nohara}, \citenamefont {Takagi},\
  and\ \citenamefont {Mason}}]{Lake-2002}%
  \BibitemOpen
  \bibfield  {author} {\bibinfo {author} {\bibfnamefont {B.}~\bibnamefont
  {Lake}}, \bibinfo {author} {\bibfnamefont {H.~M.}\ \bibnamefont {R{\o}nnow}},
  \bibinfo {author} {\bibfnamefont {N.~B.}\ \bibnamefont {Christensen}},
  \bibinfo {author} {\bibfnamefont {G.}~\bibnamefont {Aeppli}}, \bibinfo
  {author} {\bibfnamefont {K.}~\bibnamefont {Lefmann}}, \bibinfo {author}
  {\bibfnamefont {D.~F.}\ \bibnamefont {McMorrow}}, \bibinfo {author}
  {\bibfnamefont {P.}~\bibnamefont {Vorderwisch}}, \bibinfo {author}
  {\bibfnamefont {P.}~\bibnamefont {Smeibidl}}, \bibinfo {author}
  {\bibfnamefont {N.}~\bibnamefont {Mangkorntong}}, \bibinfo {author}
  {\bibfnamefont {T.}~\bibnamefont {Sasagawa}}, \bibinfo {author}
  {\bibfnamefont {M.}~\bibnamefont {Nohara}}, \bibinfo {author} {\bibfnamefont
  {H.}~\bibnamefont {Takagi}}, \ and\ \bibinfo {author} {\bibfnamefont {T.~E.}\
  \bibnamefont {Mason}},\ }\href {\doibase 10.1038/415299a} {\bibfield
  {journal} {\bibinfo  {journal} {Nature}\ }\textbf {\bibinfo {volume} {415}},\
  \bibinfo {pages} {299} (\bibinfo {year} {2002})}\BibitemShut {NoStop}%
\bibitem [{\citenamefont {Kakuyanagi}\ \emph {et~al.}(2003)\citenamefont
  {Kakuyanagi}, \citenamefont {Kumagai}, \citenamefont {Matsuda},\ and\
  \citenamefont {Hasegawa}}]{Kakuyanagi-2003}%
  \BibitemOpen
  \bibfield  {author} {\bibinfo {author} {\bibfnamefont {K.}~\bibnamefont
  {Kakuyanagi}}, \bibinfo {author} {\bibfnamefont {K.}~\bibnamefont {Kumagai}},
  \bibinfo {author} {\bibfnamefont {Y.}~\bibnamefont {Matsuda}}, \ and\
  \bibinfo {author} {\bibfnamefont {M.}~\bibnamefont {Hasegawa}},\ }\href
  {\doibase 10.1103/PhysRevLett.90.197003} {\bibfield  {journal} {\bibinfo
  {journal} {Phys. Rev. Lett.}\ }\textbf {\bibinfo {volume} {90}},\ \bibinfo
  {pages} {197003} (\bibinfo {year} {2003})}\BibitemShut {NoStop}%
\bibitem [{\citenamefont {Mounce}\ \emph {et~al.}(2011)\citenamefont {Mounce},
  \citenamefont {Oh}, \citenamefont {Mukhopadhyay}, \citenamefont {Halperin},
  \citenamefont {Reyes}, \citenamefont {Kuhns}, \citenamefont {Fujita},
  \citenamefont {Ishikado},\ and\ \citenamefont {Uchida}}]{Mounce-2011}%
  \BibitemOpen
  \bibfield  {author} {\bibinfo {author} {\bibfnamefont {A.~M.}\ \bibnamefont
  {Mounce}}, \bibinfo {author} {\bibfnamefont {S.}~\bibnamefont {Oh}}, \bibinfo
  {author} {\bibfnamefont {S.}~\bibnamefont {Mukhopadhyay}}, \bibinfo {author}
  {\bibfnamefont {W.~P.}\ \bibnamefont {Halperin}}, \bibinfo {author}
  {\bibfnamefont {A.~P.}\ \bibnamefont {Reyes}}, \bibinfo {author}
  {\bibfnamefont {P.~L.}\ \bibnamefont {Kuhns}}, \bibinfo {author}
  {\bibfnamefont {K.}~\bibnamefont {Fujita}}, \bibinfo {author} {\bibfnamefont
  {M.}~\bibnamefont {Ishikado}}, \ and\ \bibinfo {author} {\bibfnamefont
  {S.}~\bibnamefont {Uchida}},\ }\href 
  {\doibase 10.1103/PhysRevLett.106.057003} {\bibfield  {journal} {\bibinfo  {journal}
  {Phys. Rev. Lett.}\ }\textbf {\bibinfo {volume} {106}},\ \bibinfo {pages}
  {057003} (\bibinfo {year} {2011})}\BibitemShut {NoStop}%
\bibitem [{\citenamefont {Zhu}\ \emph {et~al.}(2004)\citenamefont {Zhu},
  \citenamefont {Sun}, \citenamefont {Si},\ and\ \citenamefont
  {Balatsky}}]{Zhu-2004a}%
  \BibitemOpen
  \bibfield  {author} {\bibinfo {author} {\bibfnamefont {J.-X.}\ \bibnamefont
  {Zhu}}, \bibinfo {author} {\bibfnamefont {J.}~\bibnamefont {Sun}}, \bibinfo
  {author} {\bibfnamefont {Q.}~\bibnamefont {Si}}, \ and\ \bibinfo {author}
  {\bibfnamefont {A.~V.}\ \bibnamefont {Balatsky}},\ }\href 
  {\doibase 10.1103/PhysRevLett.92.017002} {\bibfield  {journal} {\bibinfo  {journal}
  {Phys. Rev. Lett.}\ }\textbf {\bibinfo {volume} {92}},\ \bibinfo {pages}
  {017002} (\bibinfo {year} {2004})}\BibitemShut {NoStop}%
\bibitem [{\citenamefont {Bauer}\ and\ \citenamefont
  {Sachdev}(2015)}]{Bauer-2015}%
  \BibitemOpen
  \bibfield  {author} {\bibinfo {author} {\bibfnamefont {J.}~\bibnamefont
  {Bauer}}\ and\ \bibinfo {author} {\bibfnamefont {S.}~\bibnamefont
  {Sachdev}},\ }\href@noop {} {\bibfield  {journal} {\bibinfo  {journal}
  {arXiv:1506.06136}\ } (\bibinfo {year} {2015})}\BibitemShut {NoStop}%
\bibitem [{Note1()}]{Note1}%
  \BibitemOpen
  \bibinfo {note} {N. Jenkins, private communication.}\BibitemShut {Stop}%
\bibitem [{\citenamefont {Eschrig}(2006)}]{Eschrig-2006}%
  \BibitemOpen
  \bibfield  {author} {\bibinfo {author} {\bibfnamefont {M.}~\bibnamefont
  {Eschrig}},\ }\href {\doibase 10.1080/00018730600645636} {\bibfield
  {journal} {\bibinfo  {journal} {Adv. Phys.}\ }\textbf {\bibinfo {volume}
  {55}},\ \bibinfo {pages} {47} (\bibinfo {year} {2006})}\BibitemShut {NoStop}%
\bibitem [{\citenamefont {Levy~de Castro}\ \emph {et~al.}(2008)\citenamefont
  {Levy~de Castro}, \citenamefont {Berthod}, \citenamefont {Piriou},
  \citenamefont {Giannini},\ and\ \citenamefont {Fischer}}]{Levy-2008}%
  \BibitemOpen
  \bibfield  {author} {\bibinfo {author} {\bibfnamefont {G.}~\bibnamefont
  {Levy~de Castro}}, \bibinfo {author} {\bibfnamefont {C.}~\bibnamefont
  {Berthod}}, \bibinfo {author} {\bibfnamefont {A.}~\bibnamefont {Piriou}},
  \bibinfo {author} {\bibfnamefont {E.}~\bibnamefont {Giannini}}, \ and\
  \bibinfo {author} {\bibfnamefont {{\O}.}~\bibnamefont {Fischer}},\ }\href
  {\doibase 10.1103/PhysRevLett.101.267004} {\bibfield  {journal} {\bibinfo
  {journal} {Phys. Rev. Lett.}\ }\textbf {\bibinfo {volume} {101}},\ \bibinfo
  {pages} {267004} (\bibinfo {year} {2008})}\BibitemShut {NoStop}%
\bibitem [{\citenamefont {Berthod}(2010)}]{Berthod-2010}%
  \BibitemOpen
  \bibfield  {author} {\bibinfo {author} {\bibfnamefont {C.}~\bibnamefont
  {Berthod}},\ }\href {\doibase 10.1103/PhysRevB.82.024504} {\bibfield
  {journal} {\bibinfo  {journal} {Phys. Rev. B}\ }\textbf {\bibinfo {volume}
  {82}},\ \bibinfo {pages} {024504} (\bibinfo {year} {2010})}\BibitemShut
  {NoStop}%
\bibitem [{Note2()}]{Note2}%
  \BibitemOpen
  \bibinfo {note} {We use the bare hopping amplitudes and neglect a correction
  due to the magnetic field. This correction is negligible for an isolated
  vortex in the region of the core when the penetration depth is large compared
  to the core size.}\BibitemShut {Stop}%
\bibitem [{\citenamefont {Covaci}\ \emph {et~al.}(2010)\citenamefont {Covaci},
  \citenamefont {Peeters},\ and\ \citenamefont {Berciu}}]{Covaci-2010}%
  \BibitemOpen
  \bibfield  {author} {\bibinfo {author} {\bibfnamefont {L.}~\bibnamefont
  {Covaci}}, \bibinfo {author} {\bibfnamefont {F.~M.}\ \bibnamefont {Peeters}},
  \ and\ \bibinfo {author} {\bibfnamefont {M.}~\bibnamefont {Berciu}},\ }\href
  {\doibase 10.1103/PhysRevLett.105.167006} {\bibfield  {journal} {\bibinfo
  {journal} {Phys. Rev. Lett.}\ }\textbf {\bibinfo {volume} {105}},\ \bibinfo
  {pages} {167006} (\bibinfo {year} {2010})}\BibitemShut {NoStop}%
\bibitem [{\citenamefont {Soininen}\ \emph {et~al.}(1994)\citenamefont
  {Soininen}, \citenamefont {Kallin},\ and\ \citenamefont
  {Berlinsky}}]{Soininen-1994}%
  \BibitemOpen
  \bibfield  {author} {\bibinfo {author} {\bibfnamefont {P.~I.}\ \bibnamefont
  {Soininen}}, \bibinfo {author} {\bibfnamefont {C.}~\bibnamefont {Kallin}}, \
  and\ \bibinfo {author} {\bibfnamefont {A.~J.}\ \bibnamefont {Berlinsky}},\
  }\href {\doibase 10.1103/PhysRevB.50.13883} {\bibfield  {journal} {\bibinfo
  {journal} {Phys. Rev. B}\ }\textbf {\bibinfo {volume} {50}},\ \bibinfo
  {pages} {13883} (\bibinfo {year} {1994})}\BibitemShut {NoStop}%
\bibitem [{Note3()}]{Note3}%
  \BibitemOpen
  \bibinfo {note} {In our model, the vortex center sits on a lattice site. We
  have also considered a vortex centered in the middle of a plaquette as in
  Ref.~\protect \rev@citealpnum {Soininen-1994}. Both are stable
  self-consistent solutions and yield nearly identical LDOS.}\BibitemShut
  {Stop}%
\bibitem [{\citenamefont {Berthod}(2005)}]{Berthod-2005}%
  \BibitemOpen
  \bibfield  {author} {\bibinfo {author} {\bibfnamefont {C.}~\bibnamefont
  {Berthod}},\ }\href {\doibase 10.1103/PhysRevB.71.134513} {\bibfield
  {journal} {\bibinfo  {journal} {Phys. Rev. B}\ }\textbf {\bibinfo {volume}
  {71}},\ \bibinfo {pages} {134513} (\bibinfo {year} {2005})}\BibitemShut
  {NoStop}%
\bibitem [{Note4()}]{Note4}%
  \BibitemOpen
  \bibinfo {note} {A coupling $\alpha =0.7$ in our notation corresponds, in the
  notation of Ref.~\protect \rev@citealpnum {Eschrig-2000}, to $g=0.77$~eV if
  the energy-integrated susceptibility at $(\pi ,\pi )$ is $0.95\mu _{\protect
  \mathrm {B}}^2$.}\BibitemShut {Stop}%
\bibitem [{Note5()}]{Note5}%
  \BibitemOpen
  \bibinfo {note} {In the discussion of the spectral-weight transfer, we
  consider symmetric energy windows, thus ignoring subtleties associated with
  the breaking of particle-hole symmetry.}\BibitemShut {Stop}%
\bibitem [{Note6()}]{Note6}%
  \BibitemOpen
  \bibinfo {note} {See ancillary file at \url{http://arxiv.org/abs/1508.04225}
  for an animated version of Fig.~\ref {fig:fig-maps} showing data
  between $-200$ and $+200$~meV with relative and absolute color
  scales.}\BibitemShut {Stop}%
\bibitem [{\citenamefont {Volovik}(1993)}]{Volovik-1993}%
  \BibitemOpen
  \bibfield  {author} {\bibinfo {author} {\bibfnamefont {G.~E.}\ \bibnamefont
  {Volovik}},\ }\href@noop {} {\bibfield  {journal} {\bibinfo  {journal} {JETP
  Lett.}\ }\textbf {\bibinfo {volume} {58}},\ \bibinfo {pages} {469} (\bibinfo
  {year} {1993})}\BibitemShut {NoStop}%
\bibitem [{\citenamefont {Byers}\ and\ \citenamefont
  {Flatt{\'e}}(1995)}]{Byers-1995}%
  \BibitemOpen
  \bibfield  {author} {\bibinfo {author} {\bibfnamefont {J.~M.}\ \bibnamefont
  {Byers}}\ and\ \bibinfo {author} {\bibfnamefont {M.~E.}\ \bibnamefont
  {Flatt{\'e}}},\ }\href {\doibase 10.1103/PhysRevLett.74.306} {\bibfield
  {journal} {\bibinfo  {journal} {Phys. Rev. Lett.}\ }\textbf {\bibinfo
  {volume} {74}},\ \bibinfo {pages} {306} (\bibinfo {year} {1995})}\BibitemShut
  {NoStop}%
\bibitem [{\citenamefont {Niu}\ \emph {et~al.}(1995)\citenamefont {Niu},
  \citenamefont {Chang},\ and\ \citenamefont {Shih}}]{Niu-1995}%
  \BibitemOpen
  \bibfield  {author} {\bibinfo {author} {\bibfnamefont {Q.}~\bibnamefont
  {Niu}}, \bibinfo {author} {\bibfnamefont {M.~C.}\ \bibnamefont {Chang}}, \
  and\ \bibinfo {author} {\bibfnamefont {C.~K.}\ \bibnamefont {Shih}},\ }\href
  {\doibase 10.1103/PhysRevB.51.5502} {\bibfield  {journal} {\bibinfo
  {journal} {Phys. Rev. B}\ }\textbf {\bibinfo {volume} {51}},\ \bibinfo
  {pages} {5502} (\bibinfo {year} {1995})}\BibitemShut {NoStop}%
\bibitem [{\citenamefont {Xu}\ \emph {et~al.}(2006)\citenamefont {Xu},
  \citenamefont {Thibado},\ and\ \citenamefont {Ding}}]{Xu-2006}%
  \BibitemOpen
  \bibfield  {author} {\bibinfo {author} {\bibfnamefont {J.~F.}\ \bibnamefont
  {Xu}}, \bibinfo {author} {\bibfnamefont {P.~M.}\ \bibnamefont {Thibado}}, \
  and\ \bibinfo {author} {\bibfnamefont {Z.}~\bibnamefont {Ding}},\ }\href
  {\doibase 10.1063/1.2349599} {\bibfield  {journal} {\bibinfo  {journal} {Rev.
  Sci. Instrum.}\ }\textbf {\bibinfo {volume} {77}},\ \bibinfo {pages} {093703}
  (\bibinfo {year} {2006})}\BibitemShut {NoStop}%
\bibitem [{\citenamefont {Wu}\ \emph {et~al.}(2013)\citenamefont {Wu},
  \citenamefont {Mayaffre}, \citenamefont {Kr{\"a}mer}, \citenamefont
  {Horvati{\'{c}}}, \citenamefont {Berthier}, \citenamefont {Kuhns},
  \citenamefont {Reyes}, \citenamefont {Liang}, \citenamefont {Hardy},
  \citenamefont {Bonn},\ and\ \citenamefont {Julien}}]{Wu-2013}%
  \BibitemOpen
  \bibfield  {author} {\bibinfo {author} {\bibfnamefont {T.}~\bibnamefont
  {Wu}}, \bibinfo {author} {\bibfnamefont {H.}~\bibnamefont {Mayaffre}},
  \bibinfo {author} {\bibfnamefont {S.}~\bibnamefont {Kr{\"a}mer}}, \bibinfo
  {author} {\bibfnamefont {M.}~\bibnamefont {Horvati{\'{c}}}}, \bibinfo
  {author} {\bibfnamefont {C.}~\bibnamefont {Berthier}}, \bibinfo {author}
  {\bibfnamefont {P.~L.}\ \bibnamefont {Kuhns}}, \bibinfo {author}
  {\bibfnamefont {A.~P.}\ \bibnamefont {Reyes}}, \bibinfo {author}
  {\bibfnamefont {R.}~\bibnamefont {Liang}}, \bibinfo {author} {\bibfnamefont
  {W.~N.}\ \bibnamefont {Hardy}}, \bibinfo {author} {\bibfnamefont {D.~A.}\
  \bibnamefont {Bonn}}, \ and\ \bibinfo {author} {\bibfnamefont {M.-H.}\
  \bibnamefont {Julien}},\ }\href {\doibase 10.1038/ncomms3113} {\bibfield
  {journal} {\bibinfo  {journal} {Nat. Commun.}\ }\textbf {\bibinfo {volume}
  {4}},\ \bibinfo {pages} {2113} (\bibinfo {year} {2013})}\BibitemShut
  {NoStop}%
\bibitem [{\citenamefont {Inosov}(2015)}]{Inosov-2015}%
  \BibitemOpen
  \bibfield  {author} {\bibinfo {author} {\bibfnamefont {D.~S.}\ \bibnamefont
  {Inosov}},\ }\href {\doibase 10.1016/j.crhy.2015.03.001} {\bibfield
  {journal} {\bibinfo  {journal} {C. R. Physique}\ } (\bibinfo {year} {2015}),\
  10.1016/j.crhy.2015.03.001}\BibitemShut {NoStop}%
\bibitem [{\citenamefont {Shan}\ \emph {et~al.}(2012)\citenamefont {Shan},
  \citenamefont {Gong}, \citenamefont {Wang}, \citenamefont {Shen},
  \citenamefont {Hou}, \citenamefont {Ren}, \citenamefont {Li}, \citenamefont
  {Yang}, \citenamefont {Wen}, \citenamefont {Li},\ and\ \citenamefont
  {Dai}}]{Shan-2012}%
  \BibitemOpen
  \bibfield  {author} {\bibinfo {author} {\bibfnamefont {L.}~\bibnamefont
  {Shan}}, \bibinfo {author} {\bibfnamefont {J.}~\bibnamefont {Gong}}, \bibinfo
  {author} {\bibfnamefont {Y.-L.}\ \bibnamefont {Wang}}, \bibinfo {author}
  {\bibfnamefont {B.}~\bibnamefont {Shen}}, \bibinfo {author} {\bibfnamefont
  {X.}~\bibnamefont {Hou}}, \bibinfo {author} {\bibfnamefont {C.}~\bibnamefont
  {Ren}}, \bibinfo {author} {\bibfnamefont {C.}~\bibnamefont {Li}}, \bibinfo
  {author} {\bibfnamefont {H.}~\bibnamefont {Yang}}, \bibinfo {author}
  {\bibfnamefont {H.-H.}\ \bibnamefont {Wen}}, \bibinfo {author} {\bibfnamefont
  {S.}~\bibnamefont {Li}}, \ and\ \bibinfo {author} {\bibfnamefont
  {P.}~\bibnamefont {Dai}},\ }\href {\doibase 10.1103/PhysRevLett.108.227002}
  {\bibfield  {journal} {\bibinfo  {journal} {Phys. Rev. Lett.}\ }\textbf
  {\bibinfo {volume} {108}},\ \bibinfo {pages} {227002} (\bibinfo {year}
  {2012})}\BibitemShut {NoStop}%
\bibitem [{\citenamefont {Wang}\ \emph {et~al.}(2013)\citenamefont {Wang},
  \citenamefont {Yang}, \citenamefont {Fang}, \citenamefont {Shen},
  \citenamefont {Wang}, \citenamefont {Shan}, \citenamefont {Zhang},
  \citenamefont {Dai},\ and\ \citenamefont {Wen}}]{Wang-2013}%
  \BibitemOpen
  \bibfield  {author} {\bibinfo {author} {\bibfnamefont {Z.}~\bibnamefont
  {Wang}}, \bibinfo {author} {\bibfnamefont {H.}~\bibnamefont {Yang}}, \bibinfo
  {author} {\bibfnamefont {D.}~\bibnamefont {Fang}}, \bibinfo {author}
  {\bibfnamefont {B.}~\bibnamefont {Shen}}, \bibinfo {author} {\bibfnamefont
  {Q.-H.}\ \bibnamefont {Wang}}, \bibinfo {author} {\bibfnamefont
  {L.}~\bibnamefont {Shan}}, \bibinfo {author} {\bibfnamefont {C.}~\bibnamefont
  {Zhang}}, \bibinfo {author} {\bibfnamefont {P.}~\bibnamefont {Dai}}, \ and\
  \bibinfo {author} {\bibfnamefont {H.-H.}\ \bibnamefont {Wen}},\ }\href
  {\doibase 10.1038/nphys2478} {\bibfield  {journal} {\bibinfo  {journal} {Nat.
  Phys.}\ }\textbf {\bibinfo {volume} {9}},\ \bibinfo {pages} {42} (\bibinfo
  {year} {2013})}\BibitemShut {NoStop}%
\bibitem [{Note7()}]{Note7}%
  \BibitemOpen
  \bibinfo {note} {Note that the signature in the superconducting DOS of the
  coupling to a bosonic mode is a dip if the order parameter has $d_{x^2-y^2}$
  symmetry, but a break (peak in the DOS derivative) if it has $s$ symmetry;
  see Ref.~\protect \rev@citealpnum {Berthod-2010}.}\BibitemShut {Stop}%
\bibitem [{\citenamefont {Christianson}\ \emph {et~al.}(2008)\citenamefont
  {Christianson}, \citenamefont {Goremychkin}, \citenamefont {Osborn},
  \citenamefont {Rosenkranz}, \citenamefont {Lumsden}, \citenamefont
  {Malliakas}, \citenamefont {Todorov}, \citenamefont {Claus}, \citenamefont
  {Chung}, \citenamefont {Kanatzidis}, \citenamefont {Bewley},\ and\
  \citenamefont {Guidi}}]{Christianson-2008}%
  \BibitemOpen
  \bibfield  {author} {\bibinfo {author} {\bibfnamefont {A.~D.}\ \bibnamefont
  {Christianson}}, \bibinfo {author} {\bibfnamefont {E.~A.}\ \bibnamefont
  {Goremychkin}}, \bibinfo {author} {\bibfnamefont {R.}~\bibnamefont {Osborn}},
  \bibinfo {author} {\bibfnamefont {S.}~\bibnamefont {Rosenkranz}}, \bibinfo
  {author} {\bibfnamefont {M.~D.}\ \bibnamefont {Lumsden}}, \bibinfo {author}
  {\bibfnamefont {C.~D.}\ \bibnamefont {Malliakas}}, \bibinfo {author}
  {\bibfnamefont {I.~S.}\ \bibnamefont {Todorov}}, \bibinfo {author}
  {\bibfnamefont {H.}~\bibnamefont {Claus}}, \bibinfo {author} {\bibfnamefont
  {D.~Y.}\ \bibnamefont {Chung}}, \bibinfo {author} {\bibfnamefont {M.~G.}\
  \bibnamefont {Kanatzidis}}, \bibinfo {author} {\bibfnamefont {R.~I.}\
  \bibnamefont {Bewley}}, \ and\ \bibinfo {author} {\bibfnamefont
  {T.}~\bibnamefont {Guidi}},\ }\href {\doibase 10.1038/nature07625} {\bibfield
   {journal} {\bibinfo  {journal} {Nature (London)}\ }\textbf {\bibinfo
  {volume} {456}},\ \bibinfo {pages} {930} (\bibinfo {year}
  {2008})}\BibitemShut {NoStop}%
\bibitem [{\citenamefont {Weisse}\ \emph {et~al.}(2006)\citenamefont {Weisse},
  \citenamefont {Wellein}, \citenamefont {Alvermann},\ and\ \citenamefont
  {Fehske}}]{Weisse-2006}%
  \BibitemOpen
  \bibfield  {author} {\bibinfo {author} {\bibfnamefont {A.}~\bibnamefont
  {Weisse}}, \bibinfo {author} {\bibfnamefont {G.}~\bibnamefont {Wellein}},
  \bibinfo {author} {\bibfnamefont {A.}~\bibnamefont {Alvermann}}, \ and\
  \bibinfo {author} {\bibfnamefont {H.}~\bibnamefont {Fehske}},\ }\href
  {\doibase 10.1103/RevModPhys.78.275} {\bibfield  {journal} {\bibinfo
  {journal} {Rev. Mod. Phys.}\ }\textbf {\bibinfo {volume} {78}},\ \bibinfo
  {pages} {275} (\bibinfo {year} {2006})}\BibitemShut {NoStop}%
\bibitem [{\citenamefont {Todorov}\ \emph {et~al.}(1993)\citenamefont
  {Todorov}, \citenamefont {Briggs},\ and\ \citenamefont
  {Sutton}}]{Todorov-1993}%
  \BibitemOpen
  \bibfield  {author} {\bibinfo {author} {\bibfnamefont {T.~N.}\ \bibnamefont
  {Todorov}}, \bibinfo {author} {\bibfnamefont {G.~A.~D.}\ \bibnamefont
  {Briggs}}, \ and\ \bibinfo {author} {\bibfnamefont {A.~P.}\ \bibnamefont
  {Sutton}},\ }\href {\doibase 10.1088/0953-8984/5/15/010} {\bibfield
  {journal} {\bibinfo  {journal} {J. Phys.: Condens. Matter}\ }\textbf
  {\bibinfo {volume} {5}},\ \bibinfo {pages} {2389} (\bibinfo {year}
  {1993})}\BibitemShut {NoStop}%
\bibitem [{\citenamefont {Berthod}\ and\ \citenamefont
  {Giamarchi}(2011)}]{Berthod-2011}%
  \BibitemOpen
  \bibfield  {author} {\bibinfo {author} {\bibfnamefont {C.}~\bibnamefont
  {Berthod}}\ and\ \bibinfo {author} {\bibfnamefont {T.}~\bibnamefont
  {Giamarchi}},\ }\href {\doibase 10.1103/PhysRevB.84.155414} {\bibfield
  {journal} {\bibinfo  {journal} {Phys. Rev. B}\ }\textbf {\bibinfo {volume}
  {84}},\ \bibinfo {pages} {155414} (\bibinfo {year} {2011})}\BibitemShut
  {NoStop}%
\end{thebibliography}
\end{document}